\documentclass[12pt]{amsart}
\usepackage[T1]{fontenc}
\usepackage{lmodern}

\usepackage[margin=1.2in]{geometry}

\usepackage{hyperref}
\hypersetup{
  colorlinks = true,
  citecolor = blue,
  linkcolor = blue,
  urlcolor = blue
}

\usepackage{microtype}
\usepackage{subcaption}
\usepackage{pgfplots}
\usepackage{graphicx}
\usepackage{mathtools}
\usepackage{amsfonts}
\usepackage{amsmath}
\usepackage{amsaddr}
\usepackage{amsthm}
\usepackage{amssymb}
\usepackage{tikz}
\usepackage{todonotes}
\usepackage{nicefrac}
\usepackage{cleveref}
\usepackage{booktabs}
\usetikzlibrary{shapes.geometric}
\usetikzlibrary{fit}
\usetikzlibrary{arrows}
\usetikzlibrary{patterns}
\usepackage[misc]{ifsym}

\newcommand{\cnc}{\emph{Cube And Conquer}}

\usepackage{listings}
\definecolor{codegreen}{rgb}{0,0.6,0}
\definecolor{codegray}{rgb}{0.5,0.5,0.5}
\definecolor{codepurple}{rgb}{0.58,0,0.82}
\definecolor{backcolour}{rgb}{0.95,0.95,0.92}

\newcommand{\listingsttfamily}{\fontfamily{IBMPlexMono-TLF}\scriptsize}
\lstdefinestyle{mystyle}{
    backgroundcolor=\color{backcolour},   
    commentstyle=\color{codegreen},
    keywordstyle=\color{magenta},
    numberstyle=\tiny\color{codegray},
    stringstyle=\color{codepurple},
    breakatwhitespace=false,         
    breaklines=true,                 
    captionpos=b,                    
    keepspaces=true,                 
    numbers=left,                    
    numbersep=5pt,                  
    showspaces=false,                
    showstringspaces=false,
    showtabs=false,                  
    tabsize=2,
	basicstyle=\listingsttfamily
}
\DeclareMathAlphabet\EuRoman{U}{eur}{m}{n}
\SetMathAlphabet\EuRoman{bold}{U}{eur}{b}{n}
\newcommand{\euler}{\EuRoman}

\usepackage[ruled,vlined,linesnumbered,algonl]{algorithm2e}
\SetEndCharOfAlgoLine{}
\SetKwComment{Comment}{\footnotesize$\triangleright$\ }{}

\SetCommentSty{mycommfont}

\usepackage{xcolor}
\definecolor[named]{ACMBlue}{cmyk}{1,0.1,0,0.1}
\definecolor[named]{ACMYellow}{cmyk}{0,0.16,1,0}
\definecolor[named]{ACMOrange}{cmyk}{0,0.42,1,0.01}
\definecolor[named]{ACMRed}{cmyk}{0,0.90,0.86,0}   
\definecolor[named]{ACMLightRed}{cmyk}{0,0,0,0.35}
\definecolor[named]{ACMLightBlue}{cmyk}{0.49,0.01,0,0}
\definecolor[named]{ACMGreen}{cmyk}{0.20,0,1,0.19}
\definecolor[named]{ACMPurple}{cmyk}{0.55,0.6,0.1,0.15}
\definecolor[named]{ACMPurple2}{cmyk}{0.04,0.7,0.01,0.01}
\definecolor[named]{ACMPurple3}{cmyk}{0.04,0.7,0.01,0.01}
\definecolor[named]{ACMDarkBlue}{cmyk}{1,0.58,0,0.21}

\definecolor{redorange}{rgb}{0.878431, 0.235294, 0.192157}
\definecolor{lightblue}{rgb}{0.552941, 0.72549, 0.792157}
\definecolor{clearyellow}{rgb}{0.964706, 0.745098, 0}
\definecolor{midyellow}{rgb}{0.764706, 0.645098, 0.5}
\definecolor{clearorange}{rgb}{0.917647, 0.462745, 0}
\definecolor{mildgray}{rgb}{0.54902, 0.509804, 0.47451}
\definecolor{softblue}{rgb}{0.643137, 0.858824, 0.909804}
\definecolor{bluegray}{rgb}{0.141176, 0.313725, 0.603922}
\definecolor{lightgreen}{rgb}{0.709804, 0.741176, 0}
\definecolor{redpurple}{rgb}{0.835294, 0, 0.196078}
\definecolor{midblue}{rgb}{0, 0.592157, 0.662745}
\definecolor{clearpurple}{rgb}{0.67451, 0.0784314, 0.352941}
\definecolor{browngreen}{rgb}{0.333333, 0.313725, 0.145098}
\definecolor{darkestpurple}{rgb}{0.396078, 0.113725, 0.196078}
\definecolor{greypurple}{rgb}{0.294118, 0.219608, 0.298039}
\definecolor{darkturqoise}{rgb}{0, 0.239216, 0.298039}
\definecolor{darkbrown}{rgb}{0.305882, 0.211765, 0.160784}
\definecolor{midgreen}{rgb}{0.560784, 0.6, 0.243137}
\definecolor{darkred}{rgb}{0.576471, 0.152941, 0.172549}
\definecolor{darkpurple}{rgb}{0.313725, 0.027451, 0.470588}
\definecolor{darkestblue}{rgb}{0, 0.156863, 0.333333}
\definecolor{lightpurple}{rgb}{0.776471, 0.690196, 0.737255}
\definecolor{softgreen}{rgb}{0.733333, 0.772549, 0.572549}
\definecolor{offwhite}{rgb}{0.839216, 0.823529, 0.768627}
\definecolor{brightgreen}{rgb}{0.85, 0.98, 0.01}

\newcommand{\cola}{darkpurple} 
\newcommand{\colb}{midgreen!60!brightgreen} 
\newcommand{\colc}{midblue} 
\newcommand{\cold}{clearorange} 
\newcommand{\cole}{redorange} 
\newcommand{\colf}{clearyellow} 
\newcommand{\colg}{midyellow!80!redorange} 

\newcommand{\ptr}{\textsc{ptr}}

\newcommand{\step}[2]{\langle\mathsf{#1},#2\rangle}

\newcommand{\numa}{\textcolor{white}{$\euler{1}$}}
\newcommand{\numb}{\textcolor{white}{$\euler{2}$}}
\newcommand{\numc}{\textcolor{white}{$\euler{3}$}}
\newcommand{\numd}{\textcolor{white}{$\euler{4}$}}
\newcommand{\nume}{\textcolor{white}{$\euler{5}$}}
\newcommand{\numf}{\textcolor{white}{$\euler{6}$}}
\newcommand{\numg}{\textcolor{white}{$\euler{7}$}}
\newcommand{\colh}{midgreen}
\newcommand{\numh}{\textcolor{white}{$\euler{8}$}}
\newcommand{\coli}{darkturqoise!70!white}
\newcommand{\numi}{\textcolor{white}{$9$}}
\newcommand{\colj}{darkred}
\newcommand{\numj}{\textcolor{white}{$\!\euler{10}\!$}}
\newcommand{\colk}{lightpurple}
\newcommand{\numk}{\textcolor{white}{$\!\euler{11}\!$}}
\newcommand{\coll}{lightblue}
\newcommand{\numl}{\textcolor{white}{$\!\euler{12}\!$}}
\newcommand{\colm}{greypurple!70!white}
\newcommand{\numm}{\textcolor{white}{$\!\euler{13}\!$}}
\newcommand{\coln}{softgreen}
\newcommand{\numn}{\textcolor{white}{$\!\euler{14}\!$}}

\makeatletter
\def\orcidID#1{\href{http://orcid.org/#1}{\protect\raisebox{-1.25pt}{\protect\includegraphics{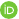}}}}
\makeatother

\newcommand{\manhattanball}[4]{
 		\foreach [evaluate=\i as \r using 2*\i]\i in {0,...,#1} {
 				\foreach \j in {0, ..., \r} {
 					\node[draw, very thick, minimum width=10mm, minimum height=10mm, fill=#4, opacity=0.5] (\i\j) at (#2 + \j - \i, #3 + #1 - \i) {};
				}		
 			}
 			\pgfmathsetmacro{\dd}{#1 - 1}
 			\foreach [evaluate=\i as \r using 2*\i]\i in {0,...,\dd} {
 				\foreach \j in {0, ..., \r} {
 					\node[draw, very thick, minimum width=10mm, minimum height=10mm, fill=#4, opacity=0.5] (\i\j) at (#2 + \j - \i, #3 - #1+\i) {};
				}		
 			}
 }
 
 \newcommand{\lowoponesquare}[5]{
 	\node[fill=#3, fill opacity=0.3, minimum width=#5, minimum height=#5, text opacity=1] (#1#2) at (#1, #2) {#4};
 }

\newcommand{\onesquare}[5]{
 	\node[fill=#3, opacity=1, minimum width=#5, minimum height=#5] (#1#2) at (#1, #2) {#4};
 }
 
 \newcommand{\dashedonesquare}[4]{
 	\node[fill=#3, opacity=1, minimum width=#4, minimum height=#4] (#1#2) at (#1, #2) {};
    \draw (#1-0.25, #2-0.25) -- (#1 + 0.25, #2 + 0.25);
    \draw (#1+0.25, #2-0.25) -- (#1 - 0.25, #2 + 0.25);
 }

\newtheorem{theorem}{Theorem}
\newtheorem{lemma}[theorem]{Lemma}
\newtheorem{definition}[theorem]{Definition}

\newtheorem{example}[theorem]{Example}

\pgfplotsset{compat=1.18}
\begin{document}
\title[The Packing Chromatic Number of the Infinite Square Grid is 15]{The Packing Chromatic Number\texorpdfstring{\\}{ }of the Infinite Square Grid is 15}
%
%
\author[Subercaseaux and Heule]{Bernardo Subercaseaux \orcidID{0000-0003-2295-1299} \and Marijn J.H.\ Heule \orcidID{0000-0002-5587-8801}}
\address{Carnegie Mellon University, Pittsburgh PA 15213, USA}
%
\email{\{bsuberca, mheule\}@cs.cmu.edu}
%
%
\begin{abstract}
A packing $k$-coloring is a natural variation on the standard notion of graph $k$-coloring, where vertices are assigned numbers from $\{1, \ldots, k\}$, and any two vertices assigned a common color $c \in \{1, \ldots, k\}$ need to be at a distance greater than $c$  (as opposed to $1$, in standard graph colorings). Despite a sequence of incremental work, determining the packing chromatic number of the infinite square grid has remained an open problem since its introduction in 2002. We culminate the search by proving this number to be 15. 
We achieve this result by improving the best-known method for this problem by roughly two orders of magnitude. The most important technique to boost performance is a novel and surprisingly effective propositional encoding. Additionally, we introduce a new symmetry-breaking approach. Since both new techniques are more complex than existing techniques for this problem, a verified approach is required to trust them. We include both techniques in a proof of unsatisfiability, reducing the trusted core to the correctness of the direct encoding. 


\end{abstract}
\maketitle     
%
%

\section{Introduction}

Automated reasoning techniques have been successfully applied to a variety of coloring problems ranging from the classical computer-assisted proof of the \emph{Four Color Theorem}~\cite{10.1215/ijm/1256049011}, to progress on the \emph{Hadwiger-Nelson problem}~\cite{Soifer2016}, or improving the bounds on Ramsey-like numbers~\cite{neiman2022}. This article contributes a new success story to the area: we show the \emph{packing} chromatic number of the infinite square grid to be 15, thus solving via automated reasoning techniques a combinatorial problem that had remained elusive for over 20 years. 

The notion of \emph{packing coloring} was introduced in the seminal work of Goddard et al.~\cite{Goddard}, and since then more than 70 articles have studied it~\cite{Brear2020}, establishing it as an active area of research. Let us consider the following definition.

\begin{definition}
A packing $k$-coloring of a simple undirected graph $G = (V, E)$ is a function~$f$ from~$V$ to $\{1, \ldots, k\}$ such that for any two distinct vertices $u, v \in V$, and any color $c \in \{1, \ldots, k\}$, it holds that $f(u)=f(v)=c$ implies $d(u, v) > c$.
\end{definition}

Note that by changing the last condition to $d(u, v) > 1$ we recover the standard notion of coloring, thus making packing colorings a natural variation of them. Intuitively, in a packing coloring, \emph{larger} colors forbid being reused in a larger region around them. Indeed, packing colorings were originally introduced under the name of \emph{broadcast coloring}, motivated by the problem of assigning broadcast frequencies to radio stations in a non-conflicting way~\cite{Goddard}, where two radio stations that are assigned the same frequency need to be at distance greater than some function of the power of their broadcast signals. Therefore, a large color represents a powerful broadcast signal at a given frequency, that cannot be reused anywhere else within a large radius around it, to avoid interference. Minimizing the number of different colors assigned can then be interpreted as minimizing the radio spectrum pollution.
The literature has preferred the name \emph{packing coloring} ever since~\cite{Brear2020}. 

Analogously to the case of standard colorings, we can naturally define the notion of \emph{packing chromatic number}, and study its computation.

\begin{definition}
	Given a graph $G = (V, E)$, define its packing chromatic number $\chi_{\rho}(G)$ as the minimum value $k$ such that $G$ admits a packing $k$-coloring.
\end{definition}

\begin{example}
	Consider the infinite graph with vertex set $\mathbb{Z}$ and with edges between consecutive integers, which we denote as $\mathbb{Z}^1$. A packing $3$-coloring is illustrated in~\Cref{fig:coloring-of-Z}. On the other hand, by examination one can observe that it is impossible to obtain a packing $2$-coloring for $\mathbb{Z}^1$. 
	
		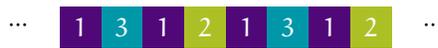
\begin{figure}[h]
		\centering
	\resizebox{.5\textwidth}{!}{
			\begin{tikzpicture}[square/.style={regular polygon,regular polygon sides=4, minimum size=40}]
			
			\onesquare{0}{0}{\cola}{\numa}{5mm}
			\onesquare{.5}{0}{\colc}{\numc}{5mm}
			\onesquare{1}{0}{\cola}{\numa}{5mm}
			\onesquare{1.5}{0}{\colb}{\numb}{5mm}
			\onesquare{2}{0}{\cola}{\numa}{5mm}
			\onesquare{2.5}{0}{\colc}{\numc}{5mm}
			\onesquare{3}{0}{\cola}{\numa}{5mm}
			\onesquare{3.5}{0}{\colb}{\numb}{5mm}
			
			\node at (-0.75,0) [] (h) {$\cdots$};
			\node at (4.25,0) [] (h) {$\cdots$};
		\end{tikzpicture}}
		\caption{Illustration of a packing $3$-coloring for $\mathbb{Z}^1$.}
		\label{fig:coloring-of-Z}
	\end{figure}
	\label{ex:Z}
\vspace{-20pt}
\end{example}

While \Cref{ex:Z} shows that $\chi_\rho(\mathbb{Z}^1) = 3$, the question of computing $\chi_\rho(\mathbb{Z}^2)$, where $\mathbb{Z}^2$ is the graph with vertex set $\mathbb{Z} \times \mathbb{Z}$ and edges between orthogonally adjacent points (i.e., points whose $\ell_1$ distance equals $1$), has been open since the introduction of packing colorings by Goddard et al.~\cite{Goddard}. On the other hand, it is known that $\chi_\rho(\mathbb{Z}^3) = \infty$ (again considering edges between points whose $\ell_1$ distance equals $1$)~\cite{FINBOW20101224}. The problem of computing 
$3 \leq \chi_\rho(\mathbb{Z}^2) \leq \infty$ has received significant attention, and it is described as ``\emph{the most attractive [of the packing coloring problems over infinite graphs]}'' by Bre\v{s}ar et al.~\cite{Brear2020}. 
We can now state our main theorem, providing a final answer to this problem.

\begin{theorem}
$\chi_\rho(\mathbb{Z}^2) = 15.$
\label{thm:main}
\end{theorem}

An upper bound of $15$ had already been proved by Martin et al.~\cite{MARTIN2017136}.
Therefore, the main contribution of our work consists of proving that $14$ colors are not enough for $\mathbb{Z}^2$. \Cref{table:historic-summary} presents a summary of the historical progress on computing $\chi_{\rho}(\mathbb{Z}^2)$.


\begin{table}
\centering
\caption{Historical summary of the bounds known for $\chi_\rho(\mathbb{Z}^2)$. }
	\begin{tabular}{p{0.7cm}lccc}
		\toprule
			Year & Citation & Approach & {\footnotesize Lower bound} & {\footnotesize Upper bound} \\ \midrule
			2002 & Goddard et al.~\cite{Goddard} & Manual & 9 & 23 \\
			2002 & Schwenk~\cite{Schwenk} & Unknown & 9 & 22\\
			2009 & Fiala et al.~\cite{10.1016/j.ejc.2008.09.014} & Manual + computer & 10 & 23\\
			2010 & Soukal and Holub~\cite{Soukal2010} & Simulated annealing & 10 & 17\\
			2010 & Ekstein et al.~\cite{DBLP:journals/corr/abs-1003-2291} & Brute force program & 12 & 17\\
			2015 & Martin et al.~\cite{martin2015packing}  & SAT solver & 13 & 16 \\
			2017 & Martin et al.~\cite{MARTIN2017136} & SAT solver & 13 & 15 \\
			2022 & Subercaseaux and Heule~\cite{subercaseaux_et_al:LIPIcs.SAT.2022.21} & SAT solver & 14 & 15 \\
			2022 & \textbf{This article}  & SAT solver & 15 & 15\\
			 \bottomrule
\end{tabular}
\label{table:historic-summary}
\end{table}

For any $k \geq 4$, the problem of determining whether a graph $G$ admits a packing $4$-coloring is known to be $\mathrm{NP}$-hard~\cite{Goddard}, and thus we do not  expect a polynomial time algorithm for computing $\chi_{\rho}(\cdot)$. This naturally motivates the use of satisfiability (SAT) solvers for studying the packing chromatic number of finite subgraphs of $\mathbb{Z}^2$. The rest of this article is thus devoted to proving~\Cref{thm:main} by using automated reasoning techniques, in a way that produces a proof that can be checked independently and that has been checked by verified software.

\section{Background}
\label{sec:background}

We start by recapitulating the components used to obtain a lower bound of 14 in our previous work~\cite{subercaseaux_et_al:LIPIcs.SAT.2022.21}.  Naturally, in order to prove a lower bound for $\mathbb{Z}^2$ one needs to prove a lower bound for a finite subgraph of it. As in earlier work, we consider \emph{disks} (i.e., $2$-dimensional balls in the $\ell_1$-metric) as the finite subgraphs to study~\cite{subercaseaux_et_al:LIPIcs.SAT.2022.21} . Concretely, let $D_r(v)$ be the subgraph induced by $\{ u \in V(\mathbb{Z}^2) \mid d(u, v) \leq r\}$. To simplify notation, we use $D_r$ as a shorthand for $D_r((0,0))$, and we let $D_{r, k}$ be the instance consisting of deciding whether $D_r$ admits a packing $k$-coloring. Moreover, let $D_{r, k, c}$ be the instance $D_{r, k}$ but enforcing that the central vertex $(0, 0)$ receives color $c$. 
\newcommand{\fsize}{110pt}

\begin{figure}[ht]
	\centering
	\resizebox{\fsize}{!}{
	\begin{tikzpicture}
			\onesquare{0}{0}{\colc}{\numc}{5mm}
			\onesquare{0}{0.5}{\cola}{\numa}{5mm}
			\onesquare{0}{2*0.5}{\colb}{\numb}{5mm}
			\onesquare{0}{3*0.5}{\cola}{\numa}{5mm}
			\onesquare{0}{-0.5}{\cola}{\numa}{5mm}
			\onesquare{0}{-2*0.5}{\colb}{\numb}{5mm}
			\onesquare{0}{-3*0.5}{\cola}{\numa}{5mm}

			\onesquare{0.5}{0}{\cola}{\numa}{5mm}
			\onesquare{0.5}{0.5}{\colg}{\numg}{5mm}
			\onesquare{0.5}{2*0.5}{\cola}{\numa}{5mm}
			\onesquare{0.5}{-0.5}{\cole}{\nume}{5mm}
			\onesquare{0.5}{-2*0.5}{\cola}{\numa}{5mm}

			\onesquare{-0.5}{0}{\cola}{\numa}{5mm}
			\onesquare{-0.5}{0.5}{\colf}{\numf}{5mm}
			\onesquare{-0.5}{2*0.5}{\cola}{\numa}{5mm}
			\onesquare{-0.5}{-0.5}{\cold}{\numd}{5mm}
			\onesquare{-0.5}{-2*0.5}{\cola}{\numa}{5mm}
			
			\onesquare{2*0.5}{0.5}{\cola}{\numa}{5mm}
			\onesquare{2*0.5}{0}{\colb}{\numb}{5mm}
			\onesquare{2*0.5}{-0.5}{\cola}{\numa}{5mm}
			
			\onesquare{-2*0.5}{0.5}{\cola}{\numa}{5mm}
			\onesquare{-2*0.5}{0}{\colb}{\numb}{5mm}
			\onesquare{-2*0.5}{-0.5}{\cola}{\numa}{5mm}
			
			\onesquare{3*0.5}{0}{\cola}{\numa}{5mm}
			
			\onesquare{-3*0.5}{0}{\cola}{\numa}{5mm}
 	\end{tikzpicture}
		}
	\hfil
	\resizebox{\fsize}{!}{
	\begin{tikzpicture}
			\onesquare{0}{0}{\colf}{\numf}{5mm}
			\onesquare{0}{0.5}{\cola}{\numa}{5mm}
			\onesquare{0}{2*0.5}{\colb}{\numb}{5mm}
			\onesquare{0}{3*0.5}{\cola}{\numa}{5mm}
			\onesquare{0}{-0.5}{\cola}{\numa}{5mm}
			\onesquare{0}{-2*0.5}{\colb}{\numb}{5mm}
			\onesquare{0}{-3*0.5}{\cola}{\numa}{5mm}

			\onesquare{0.5}{0}{\cola}{\numa}{5mm}
			\onesquare{0.5}{0.5}{\cole}{\nume}{5mm}
			\onesquare{0.5}{2*0.5}{\cola}{\numa}{5mm}
			\onesquare{0.5}{-0.5}{\colc}{\numc}{5mm}
			\onesquare{0.5}{-2*0.5}{\cola}{\numa}{5mm}

			\onesquare{-0.5}{0}{\cola}{\numa}{5mm}
			\onesquare{-0.5}{0.5}{\colc}{\numc}{5mm}
			\onesquare{-0.5}{2*0.5}{\cola}{\numa}{5mm}
			\onesquare{-0.5}{-0.5}{\cold}{\numd}{5mm}
			\onesquare{-0.5}{-2*0.5}{\cola}{\numa}{5mm}
			
			\onesquare{2*0.5}{0.5}{\cola}{\numa}{5mm}
			\onesquare{2*0.5}{0}{\colb}{\numb}{5mm}
			\onesquare{2*0.5}{-0.5}{\cola}{\numa}{5mm}
			
			\onesquare{-2*0.5}{0.5}{\cola}{\numa}{5mm}
			\onesquare{-2*0.5}{0}{\colb}{\numb}{5mm}
			\onesquare{-2*0.5}{-0.5}{\cola}{\numa}{5mm}
			
			\onesquare{3*0.5}{0}{\cola}{\numa}{5mm}
			
			\onesquare{-3*0.5}{0}{\cola}{\numa}{5mm}
 	\end{tikzpicture}
	}
	\hfil
	\resizebox{\fsize}{!}{
	\begin{tikzpicture}
			\onesquare{0}{0}{\colc}{\numc}{5mm}
			\onesquare{0}{0.5}{\cola}{\numa}{5mm}
			\onesquare{0}{2*0.5}{\colb}{\numb}{5mm}
			\onesquare{0}{3*0.5}{\cola}{\numa}{5mm}
			\onesquare{0}{-0.5}{\cola}{\numa}{5mm}
			\onesquare{0}{-2*0.5}{\colb}{\numb}{5mm}
			\onesquare{0}{-3*0.5}{\cola}{\numa}{5mm}

			\onesquare{0.5}{0}{\cola}{\numa}{5mm}
			\dashedonesquare{0.5}{0.5}{white}{5mm}
			\onesquare{0.5}{2*0.5}{\cola}{\numa}{5mm}
			\onesquare{0.5}{-0.5}{\cole}{\nume}{5mm}
			\onesquare{0.5}{-2*0.5}{\cola}{\numa}{5mm}

			\onesquare{-0.5}{0}{\cola}{\numa}{5mm}
			\onesquare{-0.5}{0.5}{\colf}{\numf}{5mm}
			\onesquare{-0.5}{2*0.5}{\cola}{\numa}{5mm}
			\onesquare{-0.5}{-0.5}{\cold}{\numd}{5mm}
			\onesquare{-0.5}{-2*0.5}{\cola}{\numa}{5mm}
			
			\onesquare{2*0.5}{0.5}{\cola}{\numa}{5mm}
			\onesquare{2*0.5}{0}{\colb}{\numb}{5mm}
			\onesquare{2*0.5}{-0.5}{\cola}{\numa}{5mm}
			
			\onesquare{-2*0.5}{0.5}{\cola}{\numa}{5mm}
			\onesquare{-2*0.5}{0}{\colb}{\numb}{5mm}
			\onesquare{-2*0.5}{-0.5}{\cola}{\numa}{5mm}
			
			\onesquare{3*0.5}{0}{\cola}{\numa}{5mm}
			
			\onesquare{-3*0.5}{0}{\cola}{\numa}{5mm}
 	\end{tikzpicture}
	}
	
	\caption{Illustration of satisfying assignments for $D_{3, 7, 3}$ and $D_{3, 6, 6}$. On the other hand, $D_{3, 6, 3}$ is not satisfiable.}
	\label{fig:d-3-7}
\end{figure}
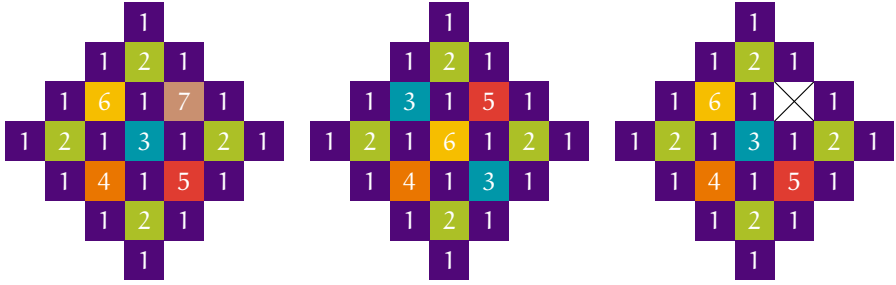

For example, a simple lemma of Subercaseaux and Heule~\cite[Proposition 5]{subercaseaux_et_al:LIPIcs.SAT.2022.21} proves that the unsatisfiability of $D_{3, 6, 3}$ is enough to deduce that $\chi_\rho(\mathbb{Z}^2) \geq 7$. We will prove a slight variation of it (\Cref{lemma:instance-to-bound}) later on in order to prove~\Cref{thm:main}, but for now let us summarize how they proved that $D_{12, 13, 12}$ is unsatisfiable.

\subsection{\bf Encodings.}

 The {\sf direct} encoding for $D_{r, k, c}$ consists simply of variables $x_{v, t}$ stating that vertex $v$ gets color $t$, as well as the following clauses:
 \begin{enumerate}
 	\item (at-least-one-color clauses, {\sc aloc}) \( \quad \quad 
 	\bigvee_{t=1}^k x_{v, t}, \quad \forall v \in V,\)
 	\item (at-most-one-distance clauses, {\sc amod}) \[
 	\overline{x_{u,t}} \lor \overline{x_{v,t}}, \quad \forall t \in \{1, \ldots, k\}, \forall u, v \in V \text{ s.t. } 0 < d(u, v) \leq t,
 	\]
 	\item (center clause) \( \quad \quad
 	x_{(0, 0), c}.
 	\)
 \end{enumerate}

This amounts to $O(r^2 k^3)$ clauses~\cite{subercaseaux_et_al:LIPIcs.SAT.2022.21}. The recursive encoding is significantly more involved, but it leads to only $O(r^2 k \log k)$ clauses asymptotically. Unfortunately, the constant involved in the asymptotic expression is large, and this encoding did not provide practical speed-ups~\cite{subercaseaux_et_al:LIPIcs.SAT.2022.21}.

\subsection{\bf Cube And Conquer.}

Introduced by Heule et al.~\cite{CnC}, the \cnc approach aims to \emph{split} a SAT instance  $\varphi$ into multiple SAT instances $\varphi_1, \ldots, \varphi_m$ in such a way that $\varphi$ is satisfiable if, and only if, at least one of the instances $\varphi_i$ is satisfiable; thus allowing to work on the different instances $\varphi_i$  in parallel. 
If $\psi = \left(c_1 \lor c_2 \lor \cdots \lor c_m\right)$ is a tautological DNF, then we have
\[
\mathrm{SAT}(\varphi) \iff \mathrm{SAT}(\varphi \land \psi) \iff \mathrm{SAT}\left( \bigvee_{i=1}^m (\varphi \land c_i)\right) \iff \mathrm{SAT}\left( \bigvee_{i=1}^m \varphi_i \right),
\]
where the different $\varphi_i \coloneqq (\varphi \land c_i)$ are the instances resulting from the split.

Intuitively, each cube $c_i$ represents a \emph{case}, i.e., an assumption about a satisfying assignment to $\varphi$, and soundness comes from $\psi$ being a tautology, which means that the split into cases is exhaustive. If the split is well designed, then each $\varphi_i$ is a particular case that is substantially easier to solve than $\varphi$, and thus solving them all in parallel can give significant speed-ups, especially considering the sequential nature of CDCL, at the core of most solvers. Our previous work~\cite{subercaseaux_et_al:LIPIcs.SAT.2022.21} proposed a concrete algorithm to generate a split, which already resulted in an almost linear speed-up, meaning that by using 128 cores, the performance gain was roughly a $\times 60$ factor.

\subsection{\bf Symmetry Breaking.}

The idea of \emph{symmetry breaking}~\cite{Crawford} consists of exploiting the symmetries that are present in SAT instances to speed-up computation. In particular, $D_{r, k, c}$ instances have 3 axes of symmetry (i.e., vertical, horizontal and diagonal) which allowed for close to an $8$-fold improvement in performance for proving $D_{12,13,12}$ to be unsatisfiable. The particular use of symmetry breaking in our previous work~\cite{subercaseaux_et_al:LIPIcs.SAT.2022.21} was happening at the \cnc level, where out of the sub-instances $\varphi_i, \ldots, \varphi_m$ produced by the split, only a $\nicefrac{1}{8}$-fraction of them had to be solved, as the rest were equivalent up to isomorphism.

\subsection{\bf Verification.}

Arguably the biggest drawback of our previous approach to prove a lower bound of 14 is that it lacked the capability of generating a computer-checkable proof. To claim a full solution to the 20-year-old problem of computing $\chi_\rho(\mathbb{Z}^2)$ that is accepted by the mathematics community, we deem paramount a fully verifiable proof that can be scrutinized independently.

The most commonly-used proofs for SAT problems are expressed in the DRAT clausal proof system~\cite{DBLP:journals/corr/Heule16}.
A DRAT proof of unsatisfiability is a list of clause addition and clause deletion steps.
Formally, a clausal proof is a list of pairs $\step{s_1}{C_1},\dots,\step{s_m}{C_m}$, where for each $i\in 1,\dots,m$, $s_i \in \{\mathsf{a}, \mathsf{d}\}$ and $C_i$ is a clause.
If $s_i = \mathsf{a}$, the pair is called an \emph{addition}, and if $s_i = \mathsf{d}$, it is called a \emph{deletion}.
For a given input formula $\varphi_0$, a clausal proof gives rise to a set of \emph{accumulated formulas} $\varphi_i$ ($i \in \{1,\dots,m\}$) as follows:
\begin{align*}
\varphi_i = 
\begin{cases}
\varphi_{i-1} \cup \{C_i\}	& \text{if $\mathsf{s}_i = \mathsf{a}$}\\
\varphi_{i-1} \setminus \{C_i\}	& \text{if $\mathsf{s}_i = \mathsf{d}$}\\
\end{cases}
\end{align*}

Each clause addition must preserve satisfiability, which is usually guaranteed by requiring the added clauses to fulfill some efficiently decidable syntactic criterion.
The main purpose of deletions is to speed up proof checking by keeping the accumulated formula small.
A valid proof of unsatisfiability must end with the addition of the empty clause.

\section{Optimizations}
\label{sec:optimizations}

Even with the best choice of parameters under the approach of Subercaseaux and Heule, solving the instance $D_{12,13,12}$ takes almost two days of computation with a $128$-core machine~\cite{subercaseaux_et_al:LIPIcs.SAT.2022.21}. In order to prove~\Cref{thm:main}, we will require to solve an instance roughly 100 times harder, and thus several optimizations will be needed.
In fact, we improve on all aspects discussed in~\Cref{sec:background} --- we present five different forms of optimization that are key to the success of our approach, which we summarize next.

\begin{enumerate}
    \item We present a new encoding, which we call the \emph{{\sf plus} encoding} that has conceptual similarities with the recursive encoding of Subercaseaux and Heule~\cite{subercaseaux_et_al:LIPIcs.SAT.2022.21}, while achieving a significant gain in practical efficiency.
    \item We present a new split algorithm that works substantially better than our previous one when coupled with the {\sf plus} encoding.
    \item We improve on symmetry breaking by using multiple layers of symmetry-breaking clauses in a way that exploits the design of the split algorithm to increase performance.
    \item We study the choice of color to fix at the center, showing that one can gain significantly in performance by making instance-based choices; for example, $D_{12, 13, 6}$ can be solved more than three times as fast as $D_{12, 13, 12}$ (the instance used in our prior work~\cite{subercaseaux_et_al:LIPIcs.SAT.2022.21}).
    \item We introduce a new and extremely simple kind of clauses called {\sc alod} clauses, which improve performance when added to the other clauses of any encoding we have tested.
\end{enumerate}
The following subsections present each of these components in detail.

\subsection{``\emph{Plus}'': a New Encoding}

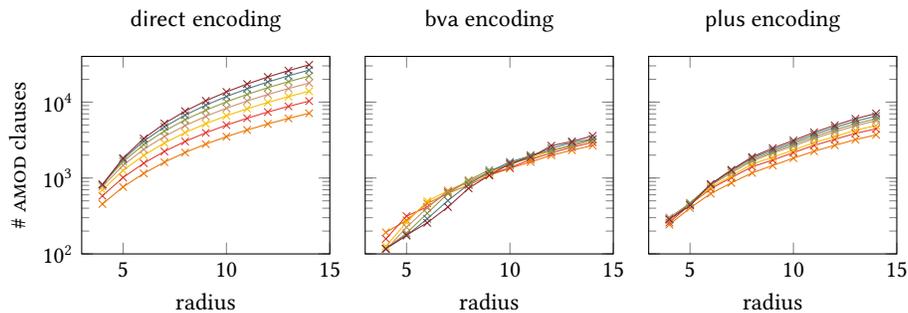
\begin{figure}
	\centering
    \begin{tikzpicture}
    	\begin{axis}[width=.32\textwidth, ylabel=$\#$ {\sc amod} clauses, ymode=log, ymin=100, ymax=40000, xlabel=radius, title={\sf direct} encoding]
    		

    \addplot[color=\cold, mark=x]
    	coordinates {
(4, 454)
(5, 762)
(6, 1150)
(7, 1618)
(8, 2166)
(9, 2794)
(10, 3502)
(11, 4290)
(12, 5158)
(13, 6106)
(14, 7134)

};

    \addplot[color=\cole, mark=x]
    	coordinates {
(4, 582)
(5, 1014)
(6, 1566)
(7, 2238)
(8, 3030)
(9, 3942)
(10, 4974)
(11, 6126)
(12, 7398)
(13, 8790)
(14, 10302)
};

\addplot[color=\colf, mark=x] coordinates {
(4, 696)
(5, 1260)
(6, 1992)
(7, 2892)
(8, 3960)
(9, 5196)
(10, 6600)
(11, 8172)
(12, 9912)
(13, 11820)
(14, 13896)
};

\addplot[color=\colg, mark=x] coordinates {
(4, 772)
(5, 1468)
(6, 2388)
(7, 3532)
(8, 4900)
(9, 6492)
(10, 8308)
(11, 10348)
(12, 12612)
(13, 15100)
(14, 17812)
};

\addplot[color=\colh, mark=x] coordinates {
(4, 820)
(5, 1644)
(6, 2756)
(7, 4156)
(8, 5844)
(9, 7820)
(10, 10084)
(11, 12636)
(12, 15476)
(13, 18604)
(14, 22020)
};

\addplot[color=\coli, mark=x] coordinates {
(4, 820)
(5, 1760)
(6, 3060)
(7, 4720)
(8, 6740)
(9, 9120)
(10, 11860)
(11, 14960)
(12, 18420)
(13, 22240)
(14, 26420)

};

\addplot[color=\colj, mark=x] coordinates {
(4, 820)
(5, 1830)
(6, 3310)
(7, 5230)
(8, 7590)
(9, 10390)
(10, 13630)
(11, 17310)
(12, 21430)
(13, 25990)
(14, 30990)
};
    	\end{axis}
    \end{tikzpicture}%
     \begin{tikzpicture}
    	\begin{axis}[width=.32\textwidth, yticklabels={,,}, ymin=100, ymax=40000, ymode=log, xlabel=radius, title={\sf bva} encoding]
    		

    \addplot[color=\cold, mark=x]
    	coordinates {
(4, 192)
(5, 277)
(6, 462)
(7, 627)
(8, 809)
(9, 1079)
(10, 1342)
(11, 1628)
(12, 1992)
(13, 2326)
(14, 2684)

};

    \addplot[color=\cole, mark=x]
    	coordinates {
(4, 158)
(5, 315)
(6, 415)
(7, 656)
(8, 902)
(9, 1117)
(10, 1367)
(11, 1751)
(12, 2140)
(13, 2541)
(14, 2957)
};

\addplot[color=\colf, mark=x] coordinates {
(4, 126)
(5, 260)
(6, 496)
(7, 678)
(8, 870)
(9, 1151)
(10, 1546)
(11, 1811)
(12, 2304)
(13, 2671)
(14, 3160)
};

\addplot[color=\colg, mark=x] coordinates {
(4, 114)
(5, 224)
(6, 404)
(7, 683)
(8, 900)
(9, 1201)
(10, 1549)
(11, 1946)
(12, 2328)
(13, 2639)
(14, 3245)
};

\addplot[color=\colh, mark=x] coordinates {
(4, 117)
(5, 188)
(6, 343)
(7, 624)
(8, 935)
(9, 1276)
(10, 1454)
(11, 2011)
(12, 2270)
(13, 2736)
(14, 3191)
};

\addplot[color=\coli, mark=x] coordinates {
(4, 117)
(5, 174)
(6, 290)
(7, 504)
(8, 824)
(9, 1213)
(10, 1627)
(11, 1959)
(12, 2462)
(13, 2929)
(14, 3271)

};

\addplot[color=\colj, mark=x] coordinates {
(4, 117)
(5, 177)
(6, 257)
(7, 415)
(8, 730)
(9, 1088)
(10, 1542)
(11, 1868)
(12, 2662)
(13, 3032)
(14, 3587)
};

    	\end{axis}

    \end{tikzpicture}%
         \begin{tikzpicture}
    	\begin{axis}[width=.32\textwidth, ymin=100, yticklabels={,,}, ymax=40000, ymode=log, xlabel=radius, title={\sf plus} encoding]
    		

    \addplot[color=\cold, mark=x]
    	coordinates {
(4, 243)
(5, 403)
(6, 623)
(7, 863)
(8, 1167)
(9, 1463)
(10, 1827)
(11, 2251)
(12, 2695)
(13, 3203)
(14, 3703)

};

    \addplot[color=\cole, mark=x]
    	coordinates {
(4, 259)
(5, 447)
(6, 723)
(7, 1011)
(8, 1407)
(9, 1723)
(10, 2163)
(11, 2691)
(12, 3231)
(13, 3879)
(14, 4447)

};

\addplot[color=\colf, mark=x] coordinates {
(4, 281)
(5, 449)
(6, 753)
(7, 1065)
(8, 1529)
(9, 1877)
(10, 2329)
(11, 2917)
(12, 3513)
(13, 4261)
(14, 4893)
};

\addplot[color=\colg, mark=x] coordinates {
(4, 303)
(5, 463)
(6, 799)
(7, 1167)
(8, 1667)
(9, 2127)
(10, 2635)
(11, 3319)
(12, 4035)
(13, 4883)
(14, 5691)
};

\addplot[color=\colh, mark=x] coordinates {
(4, 285)
(5, 469)
(6, 821)
(7, 1225)
(8, 1737)
(9, 2213)
(10, 2785)
(11, 3517)
(12, 4301)
(13, 5193)
(14, 6049)
};

\addplot[color=\coli, mark=x] coordinates {
(4, 285)
(5, 447)
(6, 831)
(7, 1271)
(8, 1831)
(9, 2331)
(10, 2939)
(11, 3751)
(12, 4619)
(13, 5607)
(14, 6535)

};

\addplot[color=\colj, mark=x] coordinates {
(4, 285)
(5, 435)
(6, 825)
(7, 1281)
(8, 1893)
(9, 2473)
(10, 3121)
(11, 3993)
(12, 4933)
(13, 6029)
(14, 7093)
};
    	\end{axis}
    \end{tikzpicture}
    
    \caption{Comparison of the size of the {\sc amod} clauses between the different encodings, for $D_4$ up to $D_{14}$ and colors $\{4, \ldots, 10\}$.}
    \label{fig:direct-vs-bva}
\end{figure}

Despite the asymptotic improvement of the recursive encoding of Subercaseaux and Heule~\cite{subercaseaux_et_al:LIPIcs.SAT.2022.21}, its contribution is mostly of ``theoretical interest'' as it does not improve solution times. Nonetheless, that encoding suggests the possibility of finding one that is both more succinct than the {\sf direct} encoding 
and that speed-ups computation. Our path towards such an encoding starts with \emph{Bounded Variable Addition (BVA)}~\cite{MHB2012}, a technique to automatically re-encode CNF formulas by adding new variables, with the goal of minimizing their resulting size (measured as the sum of the number of variables and the number of clauses). BVA can significantly reduce the size of $D_{r, k, c}$ instances, even further than the recursive encoding. Moreover, BVA actually speeds-up computation when solving the resulting instances with a CDCL solver, see~\Cref{table:bva-data}. \Cref{fig:direct-vs-bva} compares the number of {\sc amod} clauses between the {\sf direct} encoding and the BVA encoding; for example in the {\sf direct} encoding, for $D_{14}$ color $10$ would require roughly $30000$ clauses, whereas it requires roughly $3500$ in the BVA encoding. It can be observed as well in  \Cref{fig:direct-vs-bva} that the {\sf direct} encoding grows in a very structured and predictable way, where color $c$ in $D_r$ requires roughly $r^2 c^2$ clauses. On the other hand, arguably because of its locally greedy nature, the results for BVA are far more erratic, and roughly follow a $4r^2 \lg c$ curve.

The encoding resulting from BVA does not perform particularly well when coupled with the split algorithm presented in our earlier work~\cite{subercaseaux_et_al:LIPIcs.SAT.2022.21}. Indeed,~\Cref{table:bva-data} shows that while BVA heavily improves runtime under sequential CDCL, it does not provide a meaningful advantage when using \cnc. Furthermore, encodings resulting from BVA are hardly interpretable, as BVA uses a locally greedy strategy for introducing new variables. As a result, the design of a split algorithm that could work well with BVA is a very complicated task. Therefore, our approach consisted of reverse engineering what BVA was doing over some example instances, and using that insight to design a new encoding that produces instances of size comparable to those generated by BVA while being easily interpretable and thus compatible with natural split algorithms.

\begin{table}[t]
\centering
\caption{Comparison between the different encodings. \cnc~experiments follow the approach of Subercaseaux and Heule~\cite{subercaseaux_et_al:LIPIcs.SAT.2022.21} (parameters $F=5, d=2$) on a 128-core machine. Hardware details in Section~\ref{sec:experiments}. The best in each category is bolded.}
	\begin{tabular}{r@{~~}c@{~~}r@{~~~\,~~~}r@{~~}r@{~~}r@{~~}r}
		\toprule
			 & \multicolumn{2}{c}{{\sf direct} encoding} & \multicolumn{2}{c}{{\sf bva} encoding} & \multicolumn{2}{c}{{\sf plus} encoding}\\  
			 & $D_{5,10,5}$& $D_{6,11,6}$ & $D_{5,10,5}$& $D_{6,11,6}$ &$D_{5,10,5}$& $D_{6,11,6}$\\ \midrule
			{\small Number of variables}  & \phantom{00}\textbf{610} & \phantom{00}\textbf{935} & \phantom{0}973  & 1559 & \phantom{0}673 & 1039\\ 
			{\small Number of clauses}& 10688 & 21086 & \textbf{2313} & \textbf{3928} &  4063 & 7548\\ \midrule
			{\small CDCL runtime (s)} & \phantom{0}255.12 & 10774.79 & \phantom{}39.88 & 2539.38 & \phantom{00}\textbf{15.90} & \phantom{}\textbf{811.66} \\
			{\small \cnc~wall-clock (s)} & \phantom{000}0.77 & \phantom{000}26.20 & \phantom{0}0.78 & \phantom{0}17.97 & \phantom{000}\textbf{0.50} & \phantom{00}\textbf{6.68} \\
			 \bottomrule
\end{tabular}
\label{table:bva-data}
\end{table}

By manually inspecting BVA encodings one can deduce  that a fundamental part of their structure is what we call \emph{regional variables/clauses}. A regional variable $r_{S, c}$ is associated to a set of vertices $S$ and a color $c$, meaning that at least one vertex in $S$ receives color $c$. Let us illustrate their use with an example.

\begin{example}

	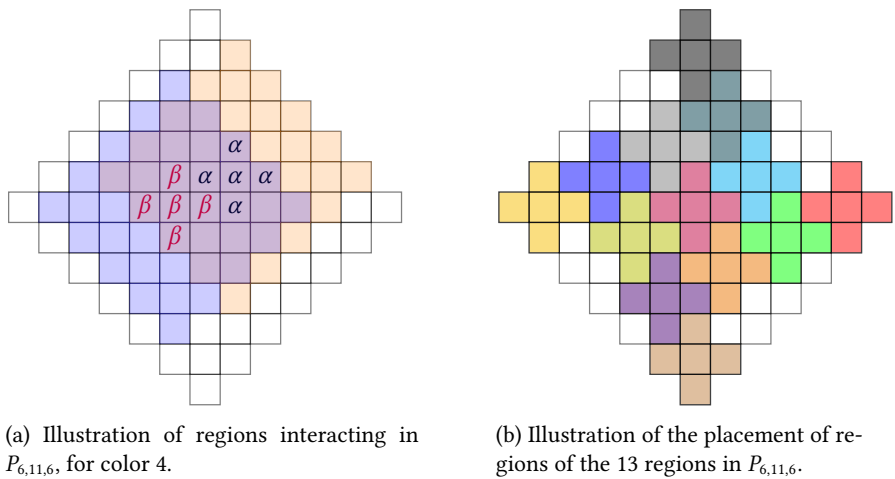
\begin{figure}[t]
\begin{subfigure}{0.45\textwidth}
	\resizebox{150pt}{150pt}{
	\begin{tikzpicture}
		\newcommand{\cone}{orange}
		\newcommand{\ctwo}{blue}
		\newcommand{\cthree}{orange!50!blue}
		\newcommand{\alphacolor}{purple}
		\newcommand{\betacolor}{black}
		\manhattanball{6}{0}{0}{white}

\lowoponesquare{4}{0}{\cone}{}{1cm}
\lowoponesquare{3}{-1}{\cone}{}{1cm}
\lowoponesquare{3}{1}{\cone}{}{1cm}
\lowoponesquare{-5}{0}{\ctwo}{}{1cm}
\lowoponesquare{5}{1}{\cone}{}{1cm}
\lowoponesquare{-3}{0}{\ctwo}{}{1cm}
\lowoponesquare{0}{2}{\cthree}{}{1cm}
\lowoponesquare{1}{-3}{\cone}{}{1cm}
\lowoponesquare{2}{2}{\cone}{}{1cm}
\lowoponesquare{1}{0}{\ctwo}{\Huge \textcolor{\alphacolor}{$\alpha$}}{1cm}
\lowoponesquare{1}{3}{\cone}{}{1cm}
\lowoponesquare{-4}{-1}{\ctwo}{}{1cm}
\lowoponesquare{-4}{1}{\ctwo}{}{1cm}
\lowoponesquare{-1}{-1}{\cone}{\Huge \textcolor{\betacolor}{$\beta$}}{1cm}
\lowoponesquare{-2}{-2}{\ctwo}{}{1cm}
\lowoponesquare{-2}{-1}{\ctwo}{}{1cm}
\lowoponesquare{-1}{-2}{\ctwo}{}{1cm}
\lowoponesquare{-2}{1}{\cthree}{}{1cm}
\lowoponesquare{-1}{1}{\cone}{\Huge \textcolor{\betacolor}{$\beta$}}{1cm}
\lowoponesquare{-1}{4}{\ctwo}{}{1cm}
\lowoponesquare{4}{2}{\cone}{}{1cm}
\lowoponesquare{3}{0}{\cthree}{}{1cm}
\lowoponesquare{3}{3}{\cone}{}{1cm}
\lowoponesquare{-3}{2}{\ctwo}{}{1cm}
\lowoponesquare{0}{-1}{\cthree}{}{1cm}
\lowoponesquare{0}{-2}{\cthree}{}{1cm}
\lowoponesquare{0}{1}{\ctwo}{\Huge \textcolor{\alphacolor}{$\alpha$}}{1cm}
\lowoponesquare{2}{-2}{\cone}{}{1cm}
\lowoponesquare{2}{-1}{\cthree}{}{1cm}
\lowoponesquare{2}{4}{\cone}{}{1cm}
\lowoponesquare{0}{4}{\cone}{}{1cm}
\lowoponesquare{1}{2}{\ctwo}{\Huge \textcolor{\alphacolor}{$\alpha$}}{1cm}
\lowoponesquare{2}{1}{\ctwo}{\Huge \textcolor{\alphacolor}{$\alpha$}}{1cm}
\lowoponesquare{1}{5}{\cone}{}{1cm}
\lowoponesquare{-4}{0}{\ctwo}{}{1cm}
\lowoponesquare{-2}{-3}{\ctwo}{}{1cm}
\lowoponesquare{-1}{-3}{\ctwo}{}{1cm}
\lowoponesquare{-2}{0}{\cone}{\Huge \textcolor{\betacolor}{$\beta$}}{1cm}
\lowoponesquare{-1}{0}{\cone}{\Huge \textcolor{\betacolor}{$\beta$}}{1cm}
\lowoponesquare{-1}{3}{\cthree}{}{1cm}
\lowoponesquare{-2}{3}{\ctwo}{}{1cm}
\lowoponesquare{3}{2}{\cone}{}{1cm}
\lowoponesquare{-1}{2}{\cthree}{}{1cm}
\lowoponesquare{4}{1}{\cone}{}{1cm}
\lowoponesquare{-3}{-2}{\ctwo}{}{1cm}
\lowoponesquare{-3}{-1}{\ctwo}{}{1cm}
\lowoponesquare{0}{-3}{\ctwo}{}{1cm}
\lowoponesquare{-3}{1}{\cthree}{}{1cm}
\lowoponesquare{0}{0}{\cone}{\Huge \textcolor{\betacolor}{$\beta$}}{1cm}
\lowoponesquare{1}{1}{\ctwo}{\Huge \textcolor{\alphacolor}{$\alpha$}}{1cm}
\lowoponesquare{0}{3}{\cthree}{}{1cm}
\lowoponesquare{2}{0}{\cthree}{}{1cm}
\lowoponesquare{1}{-2}{\cthree}{}{1cm}
\lowoponesquare{1}{-1}{\cthree}{}{1cm}
\lowoponesquare{1}{4}{\cone}{}{1cm}
\lowoponesquare{2}{3}{\cone}{}{1cm}
\lowoponesquare{-1}{-4}{\ctwo}{}{1cm}
\lowoponesquare{-2}{2}{\cthree}{}{1cm}
	\end{tikzpicture}}
		\caption{Illustration of regions interacting in $P_{6,11, 6}$, for color $5$.}
	\label{fig:ex-region-clauses}
	\end{subfigure}%
	\hspace{3em}
	\begin{subfigure}{0.4\textwidth}
\resizebox{150pt}{150pt}{
	\begin{tikzpicture}[em/.style={draw, minimum width=3mm, minimum height=3mm}]
			\manhattanball{6}{0}{0}{white}

 			\manhattanball{1}{0}{0}{purple}
 			\manhattanball{1}{1}{-2}{clearorange}
 			\manhattanball{1}{-2}{-1}{lightgreen}
 			\manhattanball{1}{2}{1}{cyan}
			\manhattanball{1}{-1}{2}{gray}
			\manhattanball{1}{-1}{-3}{darkpurple}
			\manhattanball{1}{1}{3}{darkturqoise}
			\manhattanball{1}{3}{-1}{green}
			\manhattanball{1}{-3}{1}{blue}
			\manhattanball{1}{0}{5}{black}
			\manhattanball{1}{0}{-5}{brown}
			\manhattanball{1}{5}{0}{red}
			\manhattanball{1}{-5}{0}{clearyellow}
	\end{tikzpicture}}
	\caption{Illustration of the placement of the $13$ regions in $P_{6,11, 6}$.}
	\label{fig:pluses-6-11}
	\end{subfigure}
	\caption{Illustrations for $P_{6,11, 6}$.}
\end{figure}
Consider the instance $D_{6,11}$, and let us focus on the \emph{at-most-one-distance} ({\sc amod}) clauses for color $5$. \Cref{fig:ex-region-clauses} depicts two regional clauses: an $\alpha$-region,
  whose vertices are labeled with $\alpha$, and a $\beta$-region,
   whose vertices are labeled with $\beta$. 
    Both region consist of $5$ vertices organized in a \emph{plus} ($+$) shape. We thus introduce variables $r_{\alpha, 5}$ and $r_{\beta, 5}$, defined by the following clauses:
\begin{enumerate}
	\item \(
	\overline{r_{\alpha, 5}} \lor \bigvee_{v \text{ has label } \alpha} \;  x_{v, 5},\)
	\item \(
	\overline{r_{\beta, 5}} \lor \bigvee_{v \text{ has label } \beta} \; x_{v, 5},\)
	\item \(
		r_{\alpha, 5} \lor \overline{x_{v,5}}, \; \text{ for each $v$ with label $\alpha$,}
		\)
	\item \(
		r_{\beta, 5} \lor \overline{x_{v,5}}, \; \text{ for each $v$ with label $\beta$.}
		\)
\end{enumerate}

The benefit of introducing these two new variables and $2 + (5\cdot 2) = 12$ additional clauses will be shown next, as we can use them to forbid conflicts more compactly.
Note first that each vertex $v$ participates in $|D_5(v) \cap D_6| - 1$ {\sc amod} clauses for color $5$ in the {\sf direct} encoding, where the $-1$ appears since no vertex has an {\sc amod} clause with itself. 
For the $\alpha$-vertices we have

\[
\begin{split}
|D_5((0, 1)) \cap D_6| + |D_5((1, 0)) \cap D_6|  + |D_5((1, 1)) \cap D_6|\\
+ |D_5((1, 2)) \cap D_6| + |D_5((2, 1)) \cap D_6|\\ = 61 + 61 + 55 + 50 + 50 = 277,
\end{split}
\]
and for the $\beta$-vertices we have
\[
\begin{split}
|D_5((-2, 0)) \cap D_6| + |D_5((-1, -1)) \cap D_6| + |D_5((-1, 0)) \cap D_6|\\
+ |D_5((-1, 1)) \cap D_6| + |D_5((0, 0)) \cap D_6|\\ = 50 + 55 + 61 + 55 + 61 = 282.
\end{split}
\]

This amounts to $277 + 282 - \binom{10}{2} = 514$ {\sc amod} clauses involving the labeled vertices, where the subtracted term corresponds to the clauses between labeled vertices, which otherwise would be counted twice.
 However, note that all $36$ vertices shaded in light orange or light purple (\textcolor{orange!30!white}{$\blacksquare$}, or \textcolor{blue!50!orange!30!white}{$\blacksquare$} for those in the intersection with the blue-shaded area) are at distance at most $5$ from all vertices labeled with $\alpha$, and thus each literal $r_{v, 5}$, for $v \in \left(\textcolor{orange!30!white}{\blacksquare} \cup \textcolor{blue!50!orange!30!white}{\blacksquare}\right)$, 
 is incompatible with $r_{\alpha, 5}$. This means that we can encode all conflicts between $\alpha$-vertices and $\left(\textcolor{orange!30!white}{\blacksquare} \cup \textcolor{blue!50!orange!30!white}{\blacksquare}\right)$-shaded vertices with $|\left(\textcolor{orange!30!white}{\blacksquare} \cup \textcolor{blue!50!orange!30!white}{\blacksquare}\right)| = 36$ clauses. The same can be done for $\beta$-vertices and the $36$ vertices shaded with $\left(\textcolor{blue!30!white}{\blacksquare} \cup \textcolor{blue!50!orange!30!white}{\blacksquare}\right)$. Moreover, all pairs of vertices $(x, y)$ with $x$ being an $\alpha$-vertex and $y$ being a $\beta$-vertex are in conflict, which we can encode simply with the clause $(\overline{r_{\alpha, 5}} \lor \,\overline{r_{\beta, 5}})$, instead of $5 \cdot 5 = 25$ pairwise clauses. We still need, however, to forbid that more than one $\alpha$-vertex receives color $5$, and the same for $\beta$-vertices, which can be done by simply adding all $2 \cdot \binom{5}{2} = 20$ {\sc amod} clauses between all pairs of a common label. 
 Finally, notice that not all~{\sc amod} clauses have been covered thus far, as for example the clause $\left(\overline{x_{(1, 0), 5}} \lor \overline{x_{(1, -4), 5}}\right)$ is not covered by any of the previous cases. There are $149$ such uncovered clauses we still have to add directly.
Therefore,  the total number of clauses involving $\alpha$ or $\beta$ vertices has gone down to $12 + 2 \cdot 36 + 1 + 20  + 140 = 245$ clauses, from the original $514$ clauses, by merely adding two new variables.

\label{ex:regional}
\end{example}

As shown in~\Cref{ex:regional}, the use of regional clauses can make encodings more compact, and this same idea scales even better for larger instances when the regions are larger. A key challenge for designing a \emph{regional encoding} in this manner is that it requires a choice of regions (which can even be different for every color). After trying several different strategies for defining regions, we found one that works particularly well in practice (despite not yielding an optimal number for the metric $\# \text{variables} + \#\text{clauses}$),  which we denote the \emph{{\sf plus} encoding}. The {\sf plus} encoding is based on simply using ``+'' shaped regions (i.e., $D_1$) for all colors greater than $3$, and to not introduce any changes for colors $1, 2$ and $3$ as they only amount to a very small fraction of the total size of the instances we consider. We denote with $P_{d,k,c}$ the {\sf plus} encoding of the diamond of size $d$ with $k$ colors, and the centered being colored with $c$. \Cref{fig:pluses-6-11} illustrates $P_{6,11,6}$. Interestingly, the BVA encoding opted for larger regions for the larger colors, using for example $D_2$'s or $D_3$'s as regions for color $14$. We have experimentally found this to be very ineffective when coupled with our split algorithms.
In terms of the locations of the ``+'' shaped regions, we have placed them manually through an interactive program, arriving to the conclusion that the best choice of locations consists of packing as many regions as possible and as densely around the center as possible. A more formal presentation of all the clauses involved in the {\sf plus} encoding is shown in the appendix, where the particular placement used to prove $\chi_\rho(\mathbb{Z}^2) = 15$ is presented in~\Cref{fig:plus_placement15}. Nevertheless, all its components have been illustrated in~\Cref{ex:regional}.

The exact number of clauses resulting from the {\sf plus} encoding is hard to analyze precisely, but it is clear that asymptotically it only improves from the {\sf direct} encoding by a constant multiplicative factor. \Cref{fig:direct-vs-bva} and \Cref{table:bva-data} illustrate the compactness of the {\sf plus} encoding over particular instances, and its increase in efficiency both for CDCL solving as well as with the \cnc~approach of Subercaseaux and Heule~\cite{subercaseaux_et_al:LIPIcs.SAT.2022.21}.

\subsection{Symmetry Breaking}

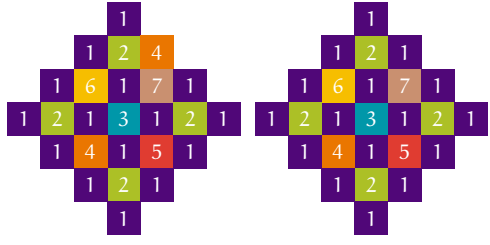
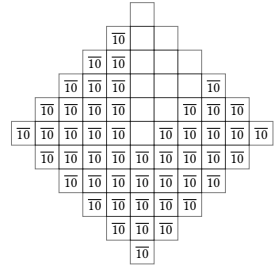
\begin{figure}[t]
	\begin{subfigure}{0.55\textwidth}
	\centering
		\resizebox{90pt}{!}{
	\begin{tikzpicture}
			\onesquare{0}{0}{\colc}{\numc}{5mm}
			\onesquare{0}{0.5}{\cola}{\numa}{5mm}
			\onesquare{0}{2*0.5}{\colb}{\numb}{5mm}
			\onesquare{0}{3*0.5}{\cola}{\numa}{5mm}
			\onesquare{0}{-0.5}{\cola}{\numa}{5mm}
			\onesquare{0}{-2*0.5}{\colb}{\numb}{5mm}
			\onesquare{0}{-3*0.5}{\cola}{\numa}{5mm}

			\onesquare{0.5}{0}{\cola}{\numa}{5mm}
			\onesquare{0.5}{0.5}{\colg}{\numg}{5mm}
			\onesquare{0.5}{2*0.5}{\cold}{\numd}{5mm}
			\onesquare{0.5}{-0.5}{\cole}{\nume}{5mm}
			\onesquare{0.5}{-2*0.5}{\cola}{\numa}{5mm}

			\onesquare{-0.5}{0}{\cola}{\numa}{5mm}
			\onesquare{-0.5}{0.5}{\colf}{\numf}{5mm}
			\onesquare{-0.5}{2*0.5}{\cola}{\numa}{5mm}
			\onesquare{-0.5}{-0.5}{\cold}{\numd}{5mm}
			\onesquare{-0.5}{-2*0.5}{\cola}{\numa}{5mm}
			
			\onesquare{2*0.5}{0.5}{\cola}{\numa}{5mm}
			\onesquare{2*0.5}{0}{\colb}{\numb}{5mm}
			\onesquare{2*0.5}{-0.5}{\cola}{\numa}{5mm}
			
			\onesquare{-2*0.5}{0.5}{\cola}{\numa}{5mm}
			\onesquare{-2*0.5}{0}{\colb}{\numb}{5mm}
			\onesquare{-2*0.5}{-0.5}{\cola}{\numa}{5mm}
			
			\onesquare{3*0.5}{0}{\cola}{\numa}{5mm}
			
			\onesquare{-3*0.5}{0}{\cola}{\numa}{5mm}
 	\end{tikzpicture}}%
 	\hfil
	\resizebox{90pt}{!}{
	\begin{tikzpicture}
			\onesquare{0}{0}{\colc}{\numc}{5mm}
			\onesquare{0}{0.5}{\cola}{\numa}{5mm}
			\onesquare{0}{2*0.5}{\colb}{\numb}{5mm}
			\onesquare{0}{3*0.5}{\cola}{\numa}{5mm}
			\onesquare{0}{-0.5}{\cola}{\numa}{5mm}
			\onesquare{0}{-2*0.5}{\colb}{\numb}{5mm}
			\onesquare{0}{-3*0.5}{\cola}{\numa}{5mm}

			\onesquare{0.5}{0}{\cola}{\numa}{5mm}
			\onesquare{0.5}{0.5}{\colg}{\numg}{5mm}
			\onesquare{0.5}{2*0.5}{\cola}{\numa}{5mm}
			\onesquare{0.5}{-0.5}{\cole}{\nume}{5mm}
			\onesquare{0.5}{-2*0.5}{\cola}{\numa}{5mm}

			\onesquare{-0.5}{0}{\cola}{\numa}{5mm}
			\onesquare{-0.5}{0.5}{\colf}{\numf}{5mm}
			\onesquare{-0.5}{2*0.5}{\cola}{\numa}{5mm}
			\onesquare{-0.5}{-0.5}{\cold}{\numd}{5mm}
			\onesquare{-0.5}{-2*0.5}{\cola}{\numa}{5mm}
			
			\onesquare{2*0.5}{0.5}{\cola}{\numa}{5mm}
			\onesquare{2*0.5}{0}{\colb}{\numb}{5mm}
			\onesquare{2*0.5}{-0.5}{\cola}{\numa}{5mm}
			
			\onesquare{-2*0.5}{0.5}{\cola}{\numa}{5mm}
			\onesquare{-2*0.5}{0}{\colb}{\numb}{5mm}
			\onesquare{-2*0.5}{-0.5}{\cola}{\numa}{5mm}
			
			\onesquare{3*0.5}{0}{\cola}{\numa}{5mm}
			
			\onesquare{-3*0.5}{0}{\cola}{\numa}{5mm}
 	\end{tikzpicture}}
 	\caption{Illustration of the effect of adding {\sc alod} clauses. The graph on the right , with {\sc alod} clauses, presents a \emph{chessboard pattern}.}
 	\label{fig:alod-clauses}
	\end{subfigure}
	\hfill
    \begin{subfigure}{0.35\textwidth}
	\resizebox{100pt}{!}{
	\begin{tikzpicture}[em/.style={draw}]
			\manhattanball{5}{0}{0}{white}
			\node at (-1,4) {\LARGE $\overline{10}$};
			\node at (-1,3) {\LARGE $\overline{10}$};
			\node at (-1,2) {\LARGE $\overline{10}$};
			\node at (-1,1) {\LARGE $\overline{10}$};
			\node at (-1,0) {\LARGE $\overline{10}$};
			\node at (-1,-1) {\LARGE $\overline{10}$};
			\node at (-1,-2) {\LARGE $\overline{10}$};
			\node at (-1,-3) {\LARGE $\overline{10}$};
			\node at (-1,-4) {\LARGE $\overline{10}$};
			\node at (-2,3) {\LARGE $\overline{10}$};
			\node at (-2,2) {\LARGE $\overline{10}$};
			\node at (-2,1) {\LARGE $\overline{10}$};
			\node at (-2,0) {\LARGE $\overline{10}$};
			\node at (-2,-1) {\LARGE $\overline{10}$};
			\node at (-2,-2) {\LARGE $\overline{10}$};
			\node at (-2,-3) {\LARGE $\overline{10}$};	
			\node at (-3,2) {\LARGE $\overline{10}$};
			\node at (-3,1) {\LARGE $\overline{10}$};
			\node at (-3,0) {\LARGE $\overline{10}$};
			\node at (-3,-1) {\LARGE $\overline{10}$};
			\node at (-3,-2) {\LARGE $\overline{10}$};	
			\node at (-4,1) {\LARGE $\overline{10}$};
			\node at (-4,0) {\LARGE $\overline{10}$};
			\node at (-4,-1) {\LARGE $\overline{10}$};							
			\node at (-5,0) {\LARGE $\overline{10}$};
			
			\node at (0,-1) {\LARGE $\overline{10}$};
			\node at (0,-2) {\LARGE $\overline{10}$};
			\node at (0,-3) {\LARGE $\overline{10}$};
			\node at (0,-4) {\LARGE $\overline{10}$};
			\node at (0,-5) {\LARGE $\overline{10}$};
			\node at (1,0) {\LARGE $\overline{10}$};
			\node at (1,-1) {\LARGE $\overline{10}$};
			\node at (1,-2) {\LARGE $\overline{10}$};
			\node at (1,-3) {\LARGE $\overline{10}$};
			\node at (1,-4) {\LARGE $\overline{10}$};
			\node at (2,1) {\LARGE $\overline{10}$};
			\node at (2,0) {\LARGE $\overline{10}$};
			\node at (2,-1) {\LARGE $\overline{10}$};
			\node at (2,-2) {\LARGE $\overline{10}$};
			\node at (2,-3) {\LARGE $\overline{10}$};
			\node at (3,2) {\LARGE $\overline{10}$};
			\node at (3,1) {\LARGE $\overline{10}$};
			\node at (3,0) {\LARGE $\overline{10}$};
			\node at (3,-1) {\LARGE $\overline{10}$};
			\node at (3,-2) {\LARGE $\overline{10}$};
			\node at (4,1) {\LARGE $\overline{10}$};
			\node at (4,0) {\LARGE $\overline{10}$};
			\node at (4,-1) {\LARGE $\overline{10}$};
			\node at (5,0) {\LARGE $\overline{10}$};

	\end{tikzpicture}}
	\caption{Some symmetry-breaking unit clauses added to $D_{5,10}$.}
	\label{fig:symmetry-breaking}
	\end{subfigure}
	\caption{The effect of adding {\sc alod} clauses (left) and symmetry-breaking (right).}
\end{figure}

Another improvement of our approach is a static symmetry-breaking technique, while Subercaseaux and Heule~\cite{subercaseaux_et_al:LIPIcs.SAT.2022.21} achieved symmetry breaking by discarding all but $\nicefrac{1}{8}$ of the cubes. We cannot do this easily since the {\sf plus} encoding does not have an $8$-fold symmetry. Instead it has a $4$-fold symmetry (see~\Cref{fig:pluses-6-11}). We add symmetry breaking clauses directly on top of the {\sf direct} encoding (i.e., instead of using it after a \cnc~split), as $D_{r, k, c}$ has indeed an $8$-fold symmetry (see~\Cref{fig:symmetry-breaking}). Concretely, if we consider a color $t$, it can only appear once in the $D_{\lfloor t/2\rfloor}$, as if appeared more than once said appearances would be at distance $\leq t$. Given this, we can assume without loss of generality that if there is one appearance of $t$ in $D_{\lfloor t/2\rfloor}$, then it appears with coordinates $(a, b)$ such that $a \geq 0 \land b \geq a$. We enforce this by adding negative units of the form $\overline{x_{(i, j), t}}$ for every pair $(i, j) \in D_{\lfloor t/2\rfloor}$ such that $i < 0 \lor j < i$. This is illustrated in~\Cref{fig:symmetry-breaking} for $D_{5, 10}$.
Note however that this can only be applied to a single color $t$, as when a vertex in the \emph{north-north-east} octant gets assigned color $t$, the $8$-fold symmetry is broken. However, if the symmetry-breaking clauses have been added for color $t$, and yet $t$ does not appear in $D_{\lfloor t/2\rfloor}$, then there is still an $8$-fold symmetry in the encoding we can exploit by breaking symmetry on some other color $t'$. This way, our encoding uses $L = 5$ \emph{layers} of symmetry breaking, for colors $k, k-1, \ldots, k-L+1$. At each layer $i$, where symmetry breaking is done over color $k-i$, except for the first (i.e., $i > 0$), we need to concatenate a clause 
\[
	{\rm SymmetryBroken}_i \coloneqq \bigvee_{t = k-i}^{k} \; \; \bigvee_{\substack{(a, b) \in D_{\lfloor t/2\rfloor} \\ 0 \leq a \leq b}} x_{(a, b), t}
\]
to each symmetry breaking clause, so that symmetry breaking is applied only when symmetry has not been broken already. \Cref{table:optimizations-6-11-6} (page 14) illustrates the impact of this symmetry breaking approach, yielding close to a $\times 40$ speed-up for $D_{6,11, 6}$.


\subsection{\emph{At-Least-One-Distance} clauses}

Yet another addition to our encoding is what we call \emph{At-Least-One-Distance} ({\sc alod}) clauses, which consist on stating that, for every vertex $v$, if we consider $D_1(v)$, then at least one vertex in $D_1(v)$ must get color $1$.  Concretely, the \emph{At-Least-One-Distance} clause corresponding to a vertex $v = (i, j)$  is 
\[
	C_v = x_{(i, j), 1} \lor x_{(i+1, j), 1} \lor x_{(i-1, j), 1} \lor x_{(i, j+1), 1} \lor x_{(i, j-1), 1}.
\]

Note that these clauses are \emph{blocked}~\cite{KULLMANN1999149} which implies that their addition preserves satisfiability. This can be seen as follows. If no vertex in $D_1(v)$ gets assigned color $1$, then we can simply assign $x_{v, 1}$, thus satisfying the new clause $C_v$.

The purpose of {\sc alod} clauses can be described as \emph{incentives} towards assigning color $1$ in a  \emph{chessboard pattern} (see \Cref{fig:alod-clauses}), which seems to simplify the rest of the computation. Empirically, their addition improves runtimes; see~\Cref{table:optimizations-6-11-6}.

\subsection{Cube And Conquer Using Auxiliary Variables}

The split of Subercaseaux and Heule~\cite{subercaseaux_et_al:LIPIcs.SAT.2022.21} is based on cases about the $x_{v, c}$ variables of the {\sf direct} encoding, and specifically using vertices $v$ that are close to the center and colors $c$ that are in the top-$t$ colors for some parameter $t$. 

Our algorithm is instead based on cases only around the new regional variables $r_{S, c}$, which appears to be key for exploiting their use in the encoding.

More concretely, our algorithm, which we call $\ptr$, is roughly based on splitting the instance into cases according to which out of the $R$ regions that are closest to the center get which of the $T$ highest colors (noting that a region can get multiple colors). A third parameter $P$ indicates the maximum number of positive literals in any cube of the split. More precisely, there are cubes with $i$ positive literals for $i \in \{0, 1, \ldots, P-1, P\}$, and the set of cubes with $i$ positive literals is constructed by $\ptr$ as follows:

\begin{enumerate}
	\item Let $\mathcal{R}$ be the set of $R$ regions that are the closest to the center, and $\mathcal{T}$ the set consisting of the $T$ highest colors (i.e., $\{k, k-1, \ldots, k-T+1 \}$).
	\item For each of the $R^i$ tuples $\vec{S}  \in \mathcal{R}^i$, we create $\binom{T}{i}$ cubes as described in the next step.
	\item For each subset $Q \subseteq \mathcal{T}$ with size $|Q| = i$, let $q_1, \ldots, q_i$ be its elements in increasing order, and then create 
	a cube with positive literals $r_{\vec{S}_j, q_j}$ for $j \in \{1, \ldots,  i\}$. Then, if $i < P$, add to the cube negative literals $\overline{r_{\vec{S}_j, q_\ell}}$ for $j \in \{1, \ldots, i\}$ and every $q_\ell \not \in Q$. 
	\end{enumerate} 

\begin{lemma}
	The cubes generated by the $\ptr$ algorithm form a tautology.
\label{lemma:split-tautology}
\end{lemma}

The proof of~\Cref{lemma:split-tautology} is quite simple, and we refer the reader to the proof of Lemma~7 in Subercaseaux and Heule~\cite{subercaseaux_et_al:LIPIcs.SAT.2022.21} for a very similar one. Moreover, because our goal is to have a verifiable proof, instead of relying on a manual proof of~\Cref{lemma:split-tautology}, we test computationally that the cubes generated by our algorithm form a tautology in all the instances mentioned in this paper.
Pseudocode for $\ptr$ is presented in the appendix as Algorithm~\ref{alg:cube_construction}.

\subsection{Optimizing the Center Color}

\begin{figure}[b!]
\centering
\begin{tikzpicture}
    \begin{axis}[
    width=.7\textwidth,height=.25\textwidth,
      scale only axis,
      xmin=1,xmax=13,
              ymode=log,
      ylabel near ticks, yticklabel pos=right,
      legend style={at={(0.90,0.90)}},
      ylabel=wall-clock time (hours),
      xtick={1,2,3,4,5,6,7,8,9,10,11,12,13},
      axis x line*=bottom]
      \addplot[color=darkpurple, mark=diamond*] coordinates {(1, 16.7111) (2, 11.644) (3, 1.56254)  (4, 1.10246)  (5, 0.347961) (6, 0.343833)  (7, 0.343433)  (8, 0.541108)  (9, 0.629939) (10, 0.795033)  (11, 0.869528)  (12, 0.870236) (13, 5.15691)};
      \legend{
              \small \textsf{wall-clock time}
            }
      
    \end{axis}
    \begin{axis}[
        width=.7\textwidth,height=.25\textwidth,
      scale only axis,
            xmin=1,xmax=13,
      ymode=log,
            legend style={at={(0.50,0.90)}},
      axis y line*=left,
      ylabel=average runtime (seconds),
      axis x line*=bottom]
      \addplot[color=redorange, mark=x] coordinates {(1, 18.1142) (2, 7.53109) (3, 2.16696) (4, 1.20158) (5, 0.509671) (6, 0.439183) (7, 0.479794) (8, 0.697709) (9, 0.899855)      (10, 1.1596) (11, 1.23483) (12, 1.17986) (13,7.37375)};
      
                  \legend{
              \small \textsf{average runtime}
            }
    \end{axis}
  \end{tikzpicture}
  
  $c$
\caption{The impact of the color in the center ($c$) on the performance for $P^\star_{12,13,c}$.}
\label{fig:center}

  \end{figure}
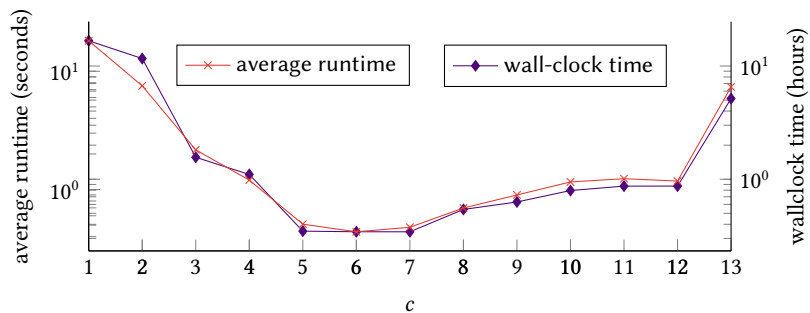

Our previous work~\cite{subercaseaux_et_al:LIPIcs.SAT.2022.21} argued that for an instance $D_{r, k}$, one should fix the color of the central vertex to $\min(r, k)$. 
However, our new experiments suggest otherwise. As the proof of~\Cref{lemma:instance-to-bound} (in the appendix) implies, we are allowed to fix any color in the center, and as long as the resulting instance is unsatisfiable, we can establish the same lower bound. 
It turns out that the choice of the center color can dramatically affect performance, as shown for instance $D_{12, 13}$ (the one used to prove $\chi_{\rho}(\mathbb{Z}^2) \geq 14$~\cite{subercaseaux_et_al:LIPIcs.SAT.2022.21}) in~\Cref{fig:center}. Interestingly, performance does not change monotonically with the value fixed in the center. 
Intuitively, it appears that fixing smaller colors in the center is ineffective as they impose restrictions on a small region around the center, while fixing very large colors in the center does not constrain the center much; for example, on the one hand, fixing a $1$ or $2$ in the center does not seem to impose any serious constraints on solutions. On the other hand, when a $12$ is fixed in the center (as in our previous work~\cite{subercaseaux_et_al:LIPIcs.SAT.2022.21}),
    color $6$ can be used $5$ times in $D_6$, whereas if color $6$ is fixed in the center, it can only be used once in $D_6$. The apparent advantage of fixing $12$ in the center (that it cannot occur anywhere else in $D_{12, 13}$), is outweighed by the extra constraints around the center that fixing color $6$ imposes; Subercaseaux and Heule had already observed that most conflicts between colors occur around the center~\cite{subercaseaux_et_al:LIPIcs.SAT.2022.21}, thus explaining why it makes sense to optimize in that area.

The main result of Subercaseaux and Heule~\cite{subercaseaux_et_al:LIPIcs.SAT.2022.21} is the unsatisfiability of $D_{12,13,12}$, which required 45 CPU hours using the same SAT solver and similar hardware. 
Let $P^\star_{d,k,c}$denote $P_{d,k,c}$ with {\sc alod} clauses and symmetry-breaking predicates.  
We show unsatisfiability of $P^\star_{12,13,12}$ in 1.18 CPU hours and of $P^\star_{12,13,6}$ in 0.34 CPU hours. So the combination 
of the {\sf plus} encoding and the improved center reduces the computational costs by two orders of magnitude.

\section{Verification} 
\label{sec:verification}

\newcommand{\dplain}{D_{15,14,6}}
\newcommand{\dsbp}{D^{\star}_{15,14,6}}
\newcommand{\psbp}{P^{\star}_{15,14,6}}
\newcommand{\taut}{N_{15,14,6}}

Our pipeline proves that, in order to trust $\chi_\rho(\mathbb{Z}^2) = 15$ as a result, the only component that requires unverified trust is the direct encoding of $\dplain$. Indeed, let $P^\star_{15, 14, 6}$ be the instance $P_{15, 14, 6}$ (with the regions indicated in~\Cref{fig:plus_placement15}), {\sc alod}-clauses, and $5$ layers of symmetry-breaking clauses, and let $\psi = \{c_1, \ldots, c_m\}$ be the set of cubes generated by the $\ptr$ algorithm with parameters $P=6, T=7, R=9$. We then prove:

\begin{enumerate}
	\item that $D_{15, 14, 6}$ is satisfiability equivalent to $P^\star_{15, 14, 6}$.
	\item the DNF $\psi = c_1 \lor c_2 \lor \cdots \lor c_m$ is a tautology.
	\item each instance $(P^\star_{15, 14, 6} \land c_i)$, for $c_i \in \psi$ is unsatisfiable. 
	\item hence the negation of each cube is implied by $P^\star_{15, 14, 6}$. 
	\item since $\psi$ is a tautology, its negation $\taut$ is unsatisfiable.
\end{enumerate}

As a result,~\Cref{thm:main} relies only on our implementation of $D_{15, 14, 6}$. Fortunately, this is quite simple, and the whole implementation is presented in Code~\ref{lst:direct-encoding} in the appendix. \Cref{fig:verification-pipeline}~illustrates the verification pipeline, and the following paragraphs detail its different components.

\begin{figure}[t!]
\[
    \mathrlap{\overbrace{\phantom{\dplain \equiv \dsbp}}^{\text{symmetry proof}}}
      \dplain \equiv 
      \mathrlap{\underbrace{\phantom{\dsbp \equiv \psbp}}_{\text{re-encoding proof}}}
      \dsbp \equiv
      \mathrlap{\overbrace{\phantom{\psbp \vDash \taut}}^{\text{implication proof}}}
      \psbp \vDash
      \underbrace{\taut \vDash \bot}_{\text{tautology proof}}
\]
\caption{Illustration of the verification pipeline.}
\label{fig:verification-pipeline}
\end{figure}

\subsection{\bf Symmetry Proof.}

The first part of the proof consists in the addition of symmetry-breaking predicates to the formula. 
This part needs to go before the re-encoding proof, because the {\sf plus} encoding  
does not have the 8-fold symmetry of the {\sf direct} encoding. Each of the clauses in the symmetry-breaking predicates have the substitution redundancy (SR) property~\cite{SR}. 
This is a very strong redundancy property and checking whether a clause $C$ has SR w.r.t. a formula $\varphi$ is NP-complete. However, since we know the symmetry, it is easy
to compute a SR certificate. There exists no SR proof checker. Instead, we implemented a prototype tool to convert SR proofs into DRAT for which formally verified checkers exists.
Our conversion is similar to the approach to converted propagation redundancy into DRAT~\cite{onevariable}. The conversion can significantly increase the size of the proof, but 
the other proof parts are typically larger for harder formulas, thus the size is acceptable. 

\subsection{\bf Re-encoding Proof.}

After symmetry breaking, the formula encoding is optimized by transforming the {\sf direct} encoding into the {\sf plus} encoding and adding the {\sc alod} clauses. 
This part of the proof is easy. All clauses in the {\sf plus} encoding, and all {\sc alod} clauses, have the RAT redundancy property w.r.t. the 
{\sf direct} encoding. This means that we can add all these clauses with a single addition step per clause. 
Afterward, the clauses that occur in the {\sf direct} encoding but not in the {\sf plus} encoding are removed using deletion steps.

\subsection{\bf Implication Proof.}

The third part of the proof expresses that the formula cannot be satisfied with any of the cubes from the split. 
For easy problems, one can avoid splitting and just use the empty cube as tautological DNF. For harder problems, splitting is crucial.
We solve $D_{15,14,6}$ using a split with just over 5 million cubes. Using a SAT solver to show that the formula
with a cube is unsatisfiable shows that the negation of the cube is implied by the formula. We can derive all these
implied clauses in parallel. The proofs of unsatisfiability can be merged into a single implication proof. 

\subsection{\bf Tautology Proof.}
The final proof part needs to show that the negation of the clauses derived in the prior steps form a tautology. 
In most cases, including ours, cubes are constructed using a tree-based method. This makes the tautology check easy
as there exists a resolution proof from the derived clauses to the empty clause using $m - 1$ resolution steps with
$m$ denoting the number of cubes. This part can be generated  using a simple SAT call. 

The final proof merges all the proof parts. In case the proof parts are all in the DRAT format, such as our proof parts,
then they can simply be merged by concatenating the proofs using the order presented above.

\section{Experiments}\label{sec:experiments}

\subsection{Experimental Setup.}
\label{subsec:details}
In terms of software, all sequential experiments were run on state-of-the-art solver {\tt CaDiCaL}~\cite{BiereFazekasFleuryHeisinger-SAT-Competition-2020-solvers}, while parallel experiments with \cnc were ran using a new implementation of parallel {\tt iCaDiCaL} because it supports incremental solving~\cite{CnC} while being significantly faster than {\tt iLingeling}.
In terms of hardware, all our experiments were run in the Bridges2~\cite{cluster} cluster from the Pittsburgh Supercomputing Center with the following specifications: Two AMD EPYC 7742 CPUs, each with 64 cores, 256MB of L3 cache, and 512GB total RAM memory.

\newcommand{\xtable}{$\times$}
\begin{table}[b]
\centering
\caption{Evaluation of the effectiveness of optimizations on $D_{6,11,6}$.}
	\begin{tabular}{ccc|rrr|rrr}
		\toprule
			sym & {\sc alod} & {\sf plus} & \#var & \#cls & {\small time (s)} & {\small derivation (MB)} & {\small proof (GB)} & {\small check (s)}\\ \midrule
  & &  & \phantom{0}935 & 21086 & 10741 & 0 & 11.99 & 31731\\
 & & \xtable  & 1039 & \phantom{0}7548 & \phantom{00}809  & 0.15 & \phantom{0}1.29 & \phantom{0}1720\\
  & \xtable & & \phantom{0}935 & 21171 & \phantom{0}8422 & 0 & \phantom{0}8.11 & 21732\\
 & \xtable & \xtable & 1039 & \phantom{0}7633 & \phantom{00}389 & 0.15 & \phantom{0}1.29 & \phantom{0}1708\\
\xtable   & & & \phantom{0}935 & 21286 & \phantom{00}273& 436 & \phantom{0}0.63  & \phantom{0}1390\\ 
 \xtable & & \xtable & 1039 & \phantom{0}7748 & \phantom{000}66 & 436 & \phantom{0}0.14 & \phantom{0}1022\\ 
  \xtable & \xtable & & \phantom{0}935 & 21371 & \phantom{00}252 & 436 & \phantom{0}0.68& \phantom{0}1359\\ 
 \xtable & \xtable & \xtable & 1039 &  \phantom{0}7833 & \phantom{000}55 & 436 & \phantom{0}0.10 & \phantom{00}997 \\  %
			 \bottomrule
\end{tabular}
\label{table:optimizations-6-11-6}
\end{table}

\subsection{\bf Effectiveness of the Optimizations.}

We evaluated the optimizations to the {\sf direct} encoding as proposed in Section~\ref{sec:optimizations}:
the {\sf plus} encoding, the addition of the {\sc alod} clauses, and the new symmetry breaking. 
The results are shown in~\Cref{table:optimizations-6-11-6}. We picked $D_{6,11,6}$ 
for this evaluation since it is the largest diamond that can still be solved within a couple of
hours on a single core. 

The main conclusion is that the optimizations significantly improve the runtime. A comparison 
between the {\sf direct} encoding without symmetry breaking and the {\sf plus} encoding with
symmetry breaking and the {\sc alod} clauses shows that the latter can be solved roughly 
200${\times}$ faster. \Cref{table:optimizations-6-11-6}~shows all 8 possible configurations. Turning on any of the optimizations
always improves performance. The effectiveness of the {\sf plus} encoding and {\sc alod} clauses
is somewhat surprising: the speed-up factor obtained by re-encoding typically does not exceed the 
factor by which the formula size is reduced. In this case, the reduction factor in formula size is less than
$3$, while the speed-up is larger than $13$ (see the difference between the first and second row of~\Cref{table:optimizations-6-11-6}). Moreover, we are not aware of the effectiveness of
adding blocked clauses, as typically SAT solvers remove them. 

We also constructed DRAT proofs of the optimizations (shown as derivation in the table) 
and the solver runtime. We merged them into a single DRAT proof by concatenating the files.
The proofs were first checked with the {\tt drat-trim} tool, which produced LRAT proofs. These LRAT
file were validated using the formally-verified {\tt cake-lpr} checker. We show the sizes of the DRAT proofs
and their corresponding checking times in~\Cref{table:optimizations-6-11-6}. Note that the checking time for the proofs
with symmetry breaking is always larger than the solving times. This is caused by expressing the 
symmetry breaking in DRAT resulting in a 436 MB proof part.


\subsection{\bf The Implication Proof.}

\begin{figure}
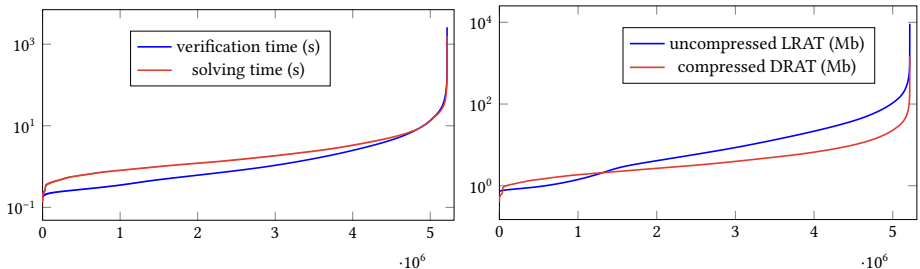

\begin{subfigure}{0.49\linewidth}

\end{subfigure}
\caption{Cactus plot of solving and verification times in seconds (left) and cactus plot of the size of the compressed DRAT proof and uncompressed LRAT proof in MB (right).}
\label{fig:cactus}
\end{figure}

The largest part of the computation consist of showing that $P^{\star}_{15,4,6}$ is unsatisfiable under each of the $5,217,031$
cubes produced by the cube generator. The results of the experiments are shown in~\Cref{fig:cactus} (left). The left plot shows that
roughly half of the cubes can be solved in a second or less. The average runtime of cubes was 3.35 seconds, while the hardest cube 
required 1584.61 seconds. 
The total runtime was 4851.38 CPU hours.

For each cube, we produced a compressed DRAT proof (the default output of {\tt CaDiCaL}). Due to the lack of hints in DRAT proofs,
they are somewhat complex to validate using a formally-verified checker. Instead, we use the tool {\tt drat-trim} to trim the proofs
and add hints. The result are uncompressed LRAT files, which we validate using the formally-verified checker {\tt cake\_lpr}.
The verification time was 4336.93 CPU hours, so slightly less than the total runtime. 

The sizes of each of the implication proofs show a similar distribution, as depicted in~\Cref{fig:cactus} (right). 
Most proofs are less than 10 MB in size. 
The compressed DRAT proofs are generally smaller
compared to the LRAT proofs, but that is mostly due to compression, which reduces the size by around 70\%.

\subsection{\bf The Chessboard Conjecture and its Counterexample.}

\begin{figure}[t]
	\centering
	\input{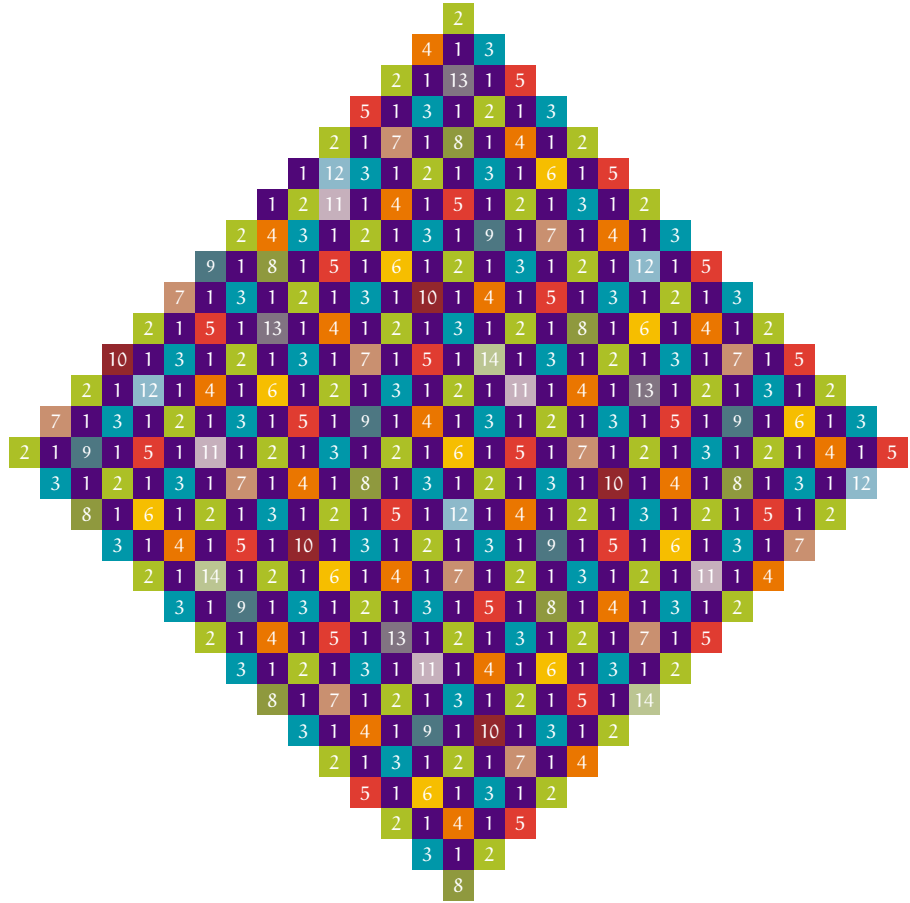}
\caption{A valid coloring of $D_{14,14,6}$. No valid coloring exists for this grid with a full chessboard pattern of 1's.}
\label{fig:chessboard-counterexample}
\end{figure}

Given that color $1$ can be used to fill in $\nicefrac{1}{2}$ of $\mathbb{Z}^2$ in a packing coloring, and the packing colorings found in the past, with $15, 16$ or $17$ colors used color $1$ with density $\nicefrac{1}{2}$ in a \emph{chessboard pattern}~\cite{MARTIN2017136}, it is tempting to assume that this must always be the case. This way, we conjectured that any instance $D_{r, k, c}$ is satisfiable if and only if it is with the chessboard pattern. The consequence of the conjecture is significant, as if it were true we could fix half of the vertices to color $1$, thus massively reducing the size of the instance and its runtime. Unfortunately, this conjecture happens to be false, with the smallest counterexample being $D_{14, 14, 6}$ as illustrated in~\Cref{fig:chessboard-counterexample}, which deviates from the chessboard pattern in only $2$ vertices. We have proved as well that no solution for $D_{14, 14, 6}$ deviating in only $1$ vertex from the chessboard pattern exists.

\paragraph{\bf Proving the Lower Bound.}

In order to prove~\Cref{thm:main}, we require the following $3$ lemmas, from where the conclusion easily follows.

\begin{lemma}
	If $D_{15, 14, 6}$ is unsatisfiable, then $\chi_\rho(\mathbb{Z}^2) \geq 15$.
	\label{lemma:instance-to-bound}
\end{lemma}%
\begin{lemma}
	If $D_{15, 14, 6}$ is satisfiable, then $P^\star_{15, 14, 6}$ is also satisfiable.
	\label{lemma:entailment-direct-new}
\end{lemma}%
\begin{lemma}
	$P^\star_{15, 14, 6}$ is unsatisfiable.
	\label{lemma:unsat-new}
\end{lemma}

We have obtained computational proofs of \Cref{lemma:entailment-direct-new} and \Cref{lemma:unsat-new} as described above, and thus it only remains to prove~\Cref{lemma:instance-to-bound}, which we include in the appendix. We can thus proceed to our main proof.

\begin{proof}[Proof of~\Cref{thm:main}]
	Since Martin et al. proved that $\chi_\rho(\mathbb{Z}^2) \leq 15$~\cite{MARTIN2017136}, it remains to show $\chi_\rho(\mathbb{Z}^2) \geq 15$, which by~\Cref{lemma:instance-to-bound} reduces to proving~\Cref{lemma:entailment-direct-new} and~\Cref{lemma:unsat-new}. We have proved these lemmas computationally, obtaining a single DRAT proof as described in~\Cref{sec:verification}.
	The total solving time was $4851.31$ CPU hours, while the total checking time of the proofs was $4336.93$ CPU hours. The total size of the compressed DRAT proof is $34$ terabytes, while the uncompressed LRAT proof weighs $122$ terabytes. 
\end{proof}


\section{Concluding Remarks and Future Work}

We have proved $\chi_\rho(\mathbb{Z}^2) = 15$ by using several SAT-solving techniques, in what constitutes a new
success story for automated reasoning tools applied to combinatorial problems. Moreover, we believe that several of our contributions in this work might be applicable to other settings and problems. Indeed, we have obtained a better encoding by reverse engineering BVA, and designed a split algorithm that works well coupled with the new encoding; this experience suggests the \emph{split-encoding compatibility} as a new key variable to pay attention to when solving combinatorial problems under the \cnc paradigm. As for future work, it is natural to study whether our techniques can be used to improve other known bounds in the packing-coloring area (see e.g.,~\cite{Brear2020}), as well as to other families of coloring problems, such as \emph{distance colorings}~\cite{KRAMER2008422}.

\vspace{0.6cm}
\paragraph{\textbf{Acknowledgments}.}
        This work is supported by the U.S.\ National Science Foundation under grant CCF-2015445. 
We thank the Pittsburgh Supercomputing Center for allowing us to use Bridges2~\cite{cluster} in our experiments.
	We thank as well the anonymous reviewers from TACAS2023 for their comments and suggestions. We also thank Donald Knuth for his thorough comments and suggestions. The first author thanks the Facebook group \emph{``actually good math problems''}, from where he first learned about this problem, and in particular to Dylan Pizzo for his post about this problem.
We are grateful as well to Zicheng Han for alerting us of an error in a previous version of~\Cref{ex:regional}.
\bibliographystyle{plain}
\bibliography{biblio}

\appendix
\section*{Appendix}

\section{Details of the Plus encoding}
Let us use notation $\Delta(v, S) := \max_{u \in S} d(u, v),$ and similarly $\Delta(S_1, S_2) := \max_{u \in S_1} \Delta(u, S_2)$.
Then, the $\textsf{plus}$ encoding $P_{r, k, c}$, with a set $R$ of variables $r_{S_i, c_i}$, consists of the following kinds of clauses:
\begin{enumerate}
	 \item (at-least-one-color clauses, {\sc aloc})
	 \[ 
 	\bigvee_{t=1}^k x_{v, t}, \quad \forall v \in V,\]
 	\item (region definition clauses)
 	\[ 
 		\overline{r_{S_i, c_i}} \lor \bigvee_{v \in S_i} x_{v, c_i}, \quad \forall r_{S_i, c_i} \in R,
 	\]
 	
 		\item (region membership clauses)
 	\[
 	r_{S_i, c_i} \lor \overline{x_{v, c_i}}, \quad \forall r_{S_i, c_i} \in R, \forall v \in S_i,
 	\]
 	
 	\item (region-vertex distance clauses)
 	\[
 	\overline{r_{S_i, c_i}}\lor \overline{x_{v, c_i}}, \quad \forall r_{S_i, c_i} \in R, \forall v \in (V \setminus S_i) \text{ s.t. } \Delta(v, S_i) \leq c_i,
 	\]
 	
 	\item (region-region distance clauses)
 	\[
 	\overline{r_{S_i, c_i}}\lor \overline{r_{S_j, c_i}}, \quad \forall r_{S_i, c_i} \in R, \forall r_{S_j, c_i} \in R\text{ s.t. } \Delta(S_i, S_j) \leq c_i \text{ and } S_i \neq S_j,
 	\]
 	
 \item (at-most-one distance clauses, {\sc amod}) \[
 	\overline{x_{u,t}} \lor \overline{x_{v,t}}, \quad \forall t \in [k], \forall u, v \in V \text{ s.t. } d(u, v) \leq t, \text { and not covered by cases 4 or 5},
 	\]

 	\item (center clause) 
 	\( \quad \quad 
 	x_{(0, 0), c}.
 	\)

\end{enumerate} 
To be even more precise about the $6^{\text{th}}$ kind of clauses, two vertices $u$ and $v$ at distance $d$ require clauses to forbid that they both take any color $t \leq d$. The $4$-th and $5$-th kind of clauses cover some of those cases, while leaving some cases uncovered. Thus, all clauses of kind $4$ or $5$ must be added first, and then clauses of kind $6$ must complete the remaining cases.

\section{Algorithm for Constructing the Cubes}

The pseudocode of~Algorithm~\ref{alg:cube_construction} is presented below. It is worth noting that in line $1.3$, the notion of \emph{closest to the center} corresponds to a distance between a set $S$ of vertices and a given vertex $u$ (i.e., the center), which is defined simply as $d(u, S) = \min_{v \in S} d(u, v).$

\vspace{0.5cm}
\begin{algorithm}[H]
\caption{CubeConstruction$(P, T, R, k, c)$; \textsc{ptr} algorithm.}\label{alg:cube_construction}
	
	$\textrm{topColors} \gets \{k, k-1, \ldots, k-T+1\}$\;
	
	\For{$i \in \emph{topColors}$}{
		$\textrm{NV}_i \gets R \text{ new variables corresponding to the regions of color } i \text{ that are closest to the center.}$\;
	}
	\If{
		$c \in \emph{topColors}$
	}{
		$\textrm{topColors} \gets \textrm{topColors} \cup \{ k- T \}$\;
		$\textrm{topColors} \gets \textrm{topColors} \setminus \{ c \}$\;
	}
	
	$ \textrm{cubes} \gets \emptyset$\;
	\For{$p \in \{P, P-1, \ldots, 0\}$}{
		\For{$\emph{colorSet} \subseteq \emph{topColors}, \, \emph{s.t. } |\emph{colorSet}| = p$} {
			
				$\textrm{products} \gets \textrm{NV}_{c_1} \times \textrm{NV}_{c_2} \times \cdots \times \textrm{NV}_{c_p
				}$, for $c_i \in \textrm{colorSet}$. \;
				$\textrm{negatives} \gets \emptyset$\;
				
				\If{$p \neq P$}{
					\For{$\emph{color} \in \emph{topColors} \setminus \emph{colorSet}$}{
						$\textrm{negatives} \gets \textrm{negatives} \cup \left(\bigcup_{S \in \textrm{NV}_{color}} \{ \overline{r_{S, \textrm{color}}}\}\right)$\;
					}
				}
				\For{$\emph{product} \in \emph{products}$}{
				$\textrm{cubes} \gets \textrm{cubes} \cup \{\{ \textrm{product} \cup \textrm{negatives} \}\}$ \;
				}
			}
		}
\end{algorithm}

\section{Code for Generating the {\sf direct} Encoding}

\renewcommand{\lstlistingname}{Code}
\begin{lstlisting}[language=Python, caption={Implementation of the {\sf direct} encoding in Python 3.10.},label={lst:direct-encoding}]
def dist(a, b):
    return abs(a[0]-b[0]) + abs(a[1]-b[1])

positions = []
for i in range(-radius, radius+1):
    for j in range(-radius, radius+1):
        if dist((i,j), (0,0)) <= radius:
            positions.append((i, j))

V = {} # map from (position, color) pairs to variable number
for pos in positions:
    for color in range(1, colors+1):
        V[(pos, color)] = len(V) + 1

clauses = []
for pos in positions:
    # each position must be assigned a color
    clauses.append([V[(pos, color)] for color in range(1, colors+1)])

for pos1 in positions:
    for pos2 in positions:
        for color in range(1, dist(pos1, pos2)+1):
            # avoid duplicate clauses
            if V[(pos1, color)] < V[(pos2, color)]:
                clauses.append([-V[(pos1, color)], -V[(pos2, color)]])

clauses.append([V[((0, 0), center_color)]])
\end{lstlisting}

\section{Proof of~\texorpdfstring{\Cref{lemma:instance-to-bound}}{Lemma 2}}


\begin{proof}[Proof of~\Cref{lemma:instance-to-bound}]
	Let us prove something slightly more general, from where the lemma follows as a particular case: that if $D_{r, k, c}$ is unsatisfiable, for some value $k$ and $\chi_\rho(\mathbb{Z}^2) \geq k$, regardless of the values of $r$ and $c \leq k$, then $\chi_\rho(\mathbb{Z}^2) \geq k+1$.
	
	Indeed, assume expecting a contradiction that $\chi_\rho(\mathbb{Z}^2) \geq k$ and  $D_{r, k, c}$ is unsatisfiable, but $\chi_\rho(\mathbb{Z}^2) \not\geq k+1$. As $\chi_\rho(\mathbb{Z}^2) \leq k$, there is a packing $k$-coloring $\varphi$ for $\mathbb{Z}^2$. Moreover, $\varphi$ must use color $k$ for at least one vertex, as we are assuming $\chi_\rho(\mathbb{Z}^2) \geq k$. There are now two cases. If $\varphi$ assigns color $c$ to some vertex $v$, then $D_{r, k, c}$ must be satisfiable, as restricting $\varphi$ to the $\ell_1$-ball of radius $r$  centered around $v$ gives us a satisfying assignment for $D_{r, k, c}$, which directly contradicts our assumption. If  $\varphi$ does not assign color $c$ to any vertex $v$, then let $\varphi'$ be equal to $\varphi$ except that $\varphi'$ assigns color $c$ to every vertex $v$ that $\varphi$ assigned color $k$. Note that $\varphi'$ is a valid packing coloring, as it only differs from $\varphi$ in that it has assigned color $c$ to vertices that had color $k$ before, and as $c \leq k$, this cannot create any conflicts, and $\varphi$ was assumed to not assign color $c$ to any vertex, so the newly colored $c$ vertices in $\varphi'$ cannot create conflicts at all. But as a result $\varphi'$ does not assign color $k$ to any vertex (note that it cannot be that $c = k$, as we have already established that $\varphi$ needs to assign color $k$ to at least one vertex given that $\chi_\rho(\mathbb{Z}^2) \geq k$), and thus $\varphi'$ is a packing $(k-1)$-coloring of $\mathbb{Z}^2$, which contradicts the assumption that $\chi_\rho(\mathbb{Z}^2) \geq k$. This concludes the proof.
\end{proof}

\newpage
\section{Plus Configuration for \texorpdfstring{$D_{15}$}{D15}}

\begin{figure}
		\centering
		\resizebox{.99\textwidth}{!}{
		\begin{tikzpicture}
			\manhattanball{15}{0}{0}{white}

\manhattanball{1}{-3}{6}{black}
\manhattanball{1}{-12}{-1}{purple}
\manhattanball{1}{6}{-2}{blue}
\manhattanball{1}{3}{-6}{black}
\manhattanball{1}{2}{6}{blue}
\manhattanball{1}{-1}{2}{cyan}
\manhattanball{1}{-2}{9}{red}
\manhattanball{1}{0}{-10}{brown}
\manhattanball{1}{-4}{3}{green}
\manhattanball{1}{9}{-3}{clearyellow}
\manhattanball{1}{-1}{-3}{gray}
\manhattanball{1}{0}{-5}{darkpurple}
\manhattanball{1}{-4}{-7}{red}
\manhattanball{1}{4}{-3}{green}
\manhattanball{1}{8}{4}{clearyellow}
\manhattanball{1}{-7}{-6}{purple}
\manhattanball{1}{6}{8}{green}
\manhattanball{1}{-11}{-3}{green}
\manhattanball{1}{3}{-11}{green}
\manhattanball{1}{7}{-4}{red}
\manhattanball{1}{-6}{-8}{green}
\manhattanball{1}{3}{-1}{gray}
\manhattanball{1}{-6}{7}{purple}
\manhattanball{1}{-11}{2}{purple}
\manhattanball{1}{3}{4}{green}
\manhattanball{1}{-13}{1}{green}
\manhattanball{1}{-10}{0}{brown}
\manhattanball{1}{-4}{-2}{darkturqoise}
\manhattanball{1}{11}{-2}{purple}
\manhattanball{1}{13}{-1}{green}
\manhattanball{1}{2}{-9}{red}
\manhattanball{1}{11}{3}{green}
\manhattanball{1}{-9}{3}{clearyellow}
\manhattanball{1}{5}{0}{darkpurple}
\manhattanball{1}{1}{-12}{purple}
\manhattanball{1}{12}{1}{purple}
\manhattanball{1}{-8}{-4}{clearyellow}
\manhattanball{1}{-3}{-9}{clearyellow}
\manhattanball{1}{-3}{-4}{green}
\manhattanball{1}{7}{6}{purple}
\manhattanball{1}{9}{2}{red}
\manhattanball{1}{-7}{4}{red}
\manhattanball{1}{5}{5}{brown}
\manhattanball{1}{6}{3}{black}
\manhattanball{1}{-7}{-1}{blue}
\manhattanball{1}{-1}{-8}{black}
\manhattanball{1}{8}{-6}{green}
\manhattanball{1}{-9}{-2}{red}
\manhattanball{1}{0}{5}{darkpurple}
\manhattanball{1}{4}{2}{darkturqoise}
\manhattanball{1}{0}{10}{brown}
\manhattanball{1}{1}{13}{green}
\manhattanball{1}{-1}{7}{blue}
\manhattanball{1}{2}{-4}{darkturqoise}
\manhattanball{1}{-2}{4}{darkturqoise}
\manhattanball{1}{1}{-7}{blue}
\manhattanball{1}{4}{7}{red}
\manhattanball{1}{-4}{8}{clearyellow}
\manhattanball{1}{-5}{5}{brown}
\manhattanball{1}{5}{-5}{brown}
\manhattanball{1}{4}{-8}{clearyellow}
\manhattanball{1}{-5}{0}{darkpurple}
\manhattanball{1}{1}{8}{black}
\manhattanball{1}{-6}{-3}{black}
\manhattanball{1}{10}{0}{brown}
\manhattanball{1}{-1}{12}{purple}
\manhattanball{1}{-2}{-6}{blue}
\manhattanball{1}{7}{1}{blue}
\manhattanball{1}{-8}{1}{black}
\manhattanball{1}{-1}{-13}{green}
\manhattanball{1}{2}{1}{cyan}
\manhattanball{1}{-2}{-1}{cyan}
\manhattanball{1}{-2}{-11}{purple}
\manhattanball{1}{8}{-1}{black}
\manhattanball{1}{-3}{1}{gray}
\manhattanball{1}{3}{9}{clearyellow}
\manhattanball{1}{-6}{2}{blue}
\manhattanball{1}{-5}{-5}{brown}
\manhattanball{1}{-8}{6}{green}
\manhattanball{1}{6}{-7}{purple}
\manhattanball{1}{1}{3}{gray}
\manhattanball{1}{-3}{11}{green}
\manhattanball{1}{0}{0}{purple}
\manhattanball{1}{1}{-2}{cyan}
\manhattanball{1}{2}{11}{purple}
		\end{tikzpicture}
		}
		\caption{Placement of the ``+'' clauses in the $D_{15}$ subgraph used for proving~\Cref{thm:main}.}\label{fig:plus_placement15}
\end{figure}

\end{document}